\documentclass[12pt, onecolumn]{IEEEtran}
\pdfoutput=1
\usepackage{float}
\usepackage{graphicx}
\usepackage{url}
\usepackage{epstopdf}
\usepackage{placeins}
\usepackage{tabularx}
\usepackage{multirow}
\usepackage{amsmath}
\usepackage{amsfonts}
\usepackage{subfig}
\usepackage{fixltx2e}
\usepackage{algorithm}
\usepackage{algcompatible}
\algnewcommand\algorithmicreturn{\textbf{return}}
\algnewcommand\RETURN{\State \algorithmicreturn}%

\begin{document}

\title{An Activity Management Algorithm for Improving Energy Efficiency of Small Cell Base Stations in 5G Heterogeneous Networks}

\author{\IEEEauthorblockN{Irmak Aykin\IEEEauthorrefmark{1} and Ezhan Karasan\IEEEauthorrefmark{2}}\\
\IEEEauthorblockA{\IEEEauthorrefmark{1} Department of Electrical and Computer Engineering, University of Arizona, USA\\ 
Email: aykin@email.arizona.edu\\
\IEEEauthorrefmark{2} Department of Electrical and Electronics Engineering, Bilkent University, Turkey\\
Email: ezhan@ee.bilkent.edu.tr}}
\maketitle

\begin{abstract}
Heterogeneous networks (HetNets) are proposed in order to meet the increasing
demand for next generation cellular wireless networks, but they also increase
the energy consumption of the base stations. In this paper, an activity management
algorithm for improving the energy efficiency of HetNets is proposed.
A smart sleep strategy is employed for the operator deployed pico base stations
to enter sleep and active modes. According to that strategy, when the number
of users exceeds the turn on threshold, the pico node becomes active and when
the number of users drop below the turn off threshold, it goes into sleep mode.
Mobile users dynamically enter and leave the cells, triggering the activation and
deactivation of pico base stations. The performance of the system is examined
for three different cellular network architectures: cell on edge (COE), uniformly
distributed cells (UDC) and macro cell only network (MoNet). Two different
user distributions are considered: uniform and hotspot. The effects of number of
hotspot users and sleep energies of pico nodes on the energy efficiency are also
investigated. The proposed activity management algorithm increases the energy
efficiency, measured in bits/J, by $20\%$. The average bit rates achieved by HetNet
users increase by $29\%$ compared with the MoNet architecture. Thus, the proposed
activity control algorithm increases the spectral efficiency of the network
while consuming the energy more efficiently.
\end{abstract}

\section*{Acknowledgments}
An earlier version of this paper has been published in the form of a Masters Thesis in July 4th, 2014.

\section{Introduction} 
\label{chp:intro}

The term Heterogeneous Network (HetNet) indicates the use of multiple types of access nodes in a wireless network. A Wide Area Network (WAN) can use macro evolved node Bs (eNBs), together with low-power eNBs such as micro, pico and femto. The large macro cells provide basic coverage, while small cells boost capacity and/or extend the range of the cellular network. Macro base stations are large cell towers that can climb as high as 75 meters and can cover an area with diameter of up to 16 kilometers; whereas, micro cells are generally employed in suburban areas and provide a coverage diameter of less than a mile. On the other hand, pico cells cover areas of a diameter around 225 meters and are typically used for indoor applications; whereas, femto base stations are about the size of a Wi-Fi router and can typically support 2 to 4 simultaneous mobile phone calls \cite{karsi}. Heterogeneous networks are seen as a promising way to meet the increasing demand for mobile broadband traffic in next generation cellular networks \cite{whitepaper}. That is why, when a 2G or 3G deployment typically consists of multisector macro base stations only, with 4G and 5G cellular networks using LTE Advanced, HetNet deployment model which consists of small cells with micro, pico and femto base stations on top of macro base station will be used \cite{figs}.

However, with every eNB turning on, we encounter fixed offset power independent from the number of users being served, and that offset power is comparable to, or even larger than, the load dependent power. That is why, activating small cells all the time, for a little number of users, is energy inefficient. We can increase the spectral efficiency as the number of active small cells increase, but then, we may consume an excessive amount of power. According to \cite{gh4}, in 2020, the population of small cells is estimated to be around 100 million with 500 million mobile users. The power consumption of a small cell today is around 6 - 10 W, and it is assumed that a small cell in 2020 will still consume approximately 5 W. Then, 100 million small cells in 2020 will consume approximately 4.4 TWh. Thus, it is crucial for small cell base stations to have some kind of a sleep strategy. If a power control mechanism is employed, the cost savings corresponding to the energy savings of HetNets are expected to be \$1.4 - \$1.6 billion in 2020. In addition, we can significantly reduce the CO\textsubscript{2} emission, using a power control mechanism \cite{greenhet2}. 

There are several suggestions to improve the energy efficiency of heterogeneous networks. Previous work in these areas can be briefly summarized as follows: \cite{greenhet2} proposes an energy-efficient deployment of the cells where the small cell base stations are arranged around the edge of the reference macro cell, and the deployment is referred to as cell-on-edge (COE) deployment. The proposed deployment ensures an increase in the network spectral and energy efficiency by facilitating cell edge mobile users with small cells. However, entirely coordinated small cell deployment is not a realistic scenario, and also COE does not improve the energy efficiency for mobile users with random movement. \cite{greenhet1} introduces active/sleep (on/off) modes in macro cell base stations and investigate the performance. However, it is risky to turn on/off macro eNB, since it provides the main coverage. \cite{greenhet6} shows that the energy efficiency of the two-tier networks with orthogonal spectrum deployment is better than that with co-channel spectrum deployment; thus, it is better not to allocate the same frequency to small and macro cells. \cite{yeni1} suggests switching off unnecessary cells to adapt actual traffic demand, but do not give details of its implementation. On the other hand, \cite{sleep} introduces energy-efficient sleep mode algorithms for small cell base stations in a bid to reduce cellular networks' power consumption. It gives detailed explanations of several sleep mode algorithms, but do not compare their performances in detail. \cite{yeni2} also introduces two node sleep modes operating on a fast and intermediate time scale respectively, in order to exploit short and longer idle periods of the nodes. However, the sleep modes it introduces are Micro DTX and Pico sleep; thus, it basically suggests different types of sleep modes for different types of small cells.

In this paper, a smart sleep strategy is employed for energy efficiency and two thresholds for the operator deployed pico node to enter sleep and active modes are introduced. When the number of users exceeds the turn on threshold, the pico node becomes active and when the number of users drop below the turn off threshold, it goes into sleep mode again. Mobile users dynamically enter and leave the cells, triggering the activation and deactivation of pico base stations. The thresholds should be sufficiently far. Otherwise, when they are too close, the base station may oscillate between two modes, on and off. And since it cannot serve a user right after turning on, this oscillation causes delay and wasted energy. Thus, the proposed activity management algorithm has hysteresis behaviour.

On the other hand, when the users' motions are not random, i.e., they move according to a hotspot model so as to enter and leave pico cells in groups, HetNet becomes more beneficial. Because, in that model, the number of users per pico cell increases and in return, the effect of offset power of the base station decreases. Also, in that case, only one threshold is enough, since the pico-eNBs do not turn on and off quite often and probability of oscillation is very low. In return, second threshold does not significantly change the performance of the system.

In this paper, the energy-efficient small cell deployment model in \cite{greenhet2} is used, together with other well-known deployment models. In addition to \cite{greenhet2}, the performance of this deployment is inspected for the case with mobile hotspot users. Also, similar to \cite{yeni1}, an algorithm that allows unnecessary small cells to turn off is used. However, different from \cite{yeni1}, this algorithm is described in detail and its performance is evaluated. Moreover, a threshold model similar to \cite{yeni2} is introduced. However, the threshold model in \cite{yeni2} is based on the traffic, where our thresholds are in terms of number of users. Also, different from \cite{yeni2}, we examine the effect of number of hotspot users and sleep energies of pico nodes on energy efficiency. According to our simulation results, when using our HetNet models, we are able to increase the energy efficiency in terms of $ bits/J $ by \%20. We can also provide a bit rate up to $ 5.88\times10^{5} $ b/s to HetNet users where the users served by the macro cell only network (MoNet) can get at most $ 5.28\times10^{5} $ b/s. In the case of hotspot users, the bit rate provided by HetNet can get as high as $ 6.8\times10^{5} $ b/s.

The rest of this paper is organized as follows. In Section \ref{chp:intro}, the related background information is provided. The differences between wired and wireless communication is explained. The evolutions of cellular networks and heterogeneous networks are described. In addition, wireless communication properties are explained and related network configurations, base station power consumption models, user movement models and distributions are outlined. In Section \ref{system}, the system model is described in detail. In Section \ref{chp:proposed}, a scheduling problem with the goal of maximizing the overall $ bits/J $ is introduced and how the problem fits into the 5G context is explained. Then, the proposed activity management algorithm and the employed hotspot model are introduced. In Section \ref{chp:simulation}, simulation environment including network topology, user distributions, mobility, activity, energy and channel models is explained. Then the simulation results are provided with detailed explanations. Finally, in Section \ref{chp:conc}, the problem and proposed solutions are summarized and the deductions of the results are made.

\subsection{Background Information}
\label{chp:background}
\subsubsection{Wireless vs. Wired Communications}

Wireless communication is the information exchange between two or more devices without using conducting wires \cite{12}. Most wireless systems, such as two-way radios and cellular telephones, use radio waves for communication. Wired communication, on the other hand, is the data transmission with the use of wires and cables. Today's television transmission is an example of wired communication. The main advantage of wireless systems is their flexibility, i.e., they allow users to be mobile during communication. However, there are two aspects of wireless communication that make it a challenging problem. First one is fading, which is the attenuation affecting a signal over certain propagation media. This attenuation may either be due to the small-scale effect of multipath fading, as well as larger-scale effects such as path loss and shadowing by obstacles. Second, there is considerable interference between wireless users, since they share the same medium, unlike wired users that have their own isolated links. The interference can be between transmitters communicating with the same receiver (e.g., uplink of a cellular system), between signals from a single transmitter to multiple receivers (e.g., downlink of a cellular system), or between different transmitter–receiver pairs (e.g., interference between users in different cells) \cite{funds}.

\subsubsection{Cellular Networks}

\begin{figure}[thb]
\begin{center}
 \centerline{%
 \resizebox{0.5\textwidth}{!}{\includegraphics{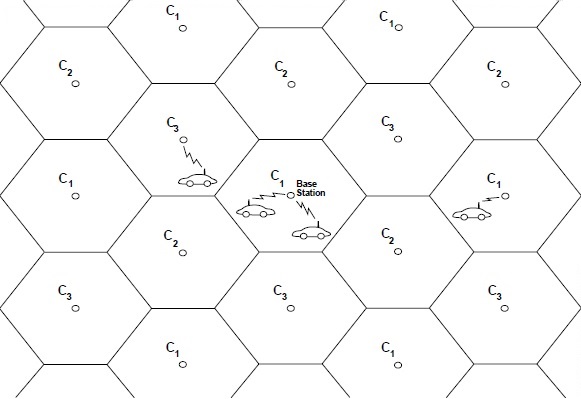}}%
 }
\caption{Cellular Systems \cite{andrea}}

\label{fig:cell}
\end{center}
\end{figure}

Today, perhaps the most commonly used wireless systems are cellular networks. A cellular network consists of a large number of wireless subscribers with cellular telephones, which can be used in cars, in buildings, on the street, or almost anywhere. There are also several fixed base-stations, which are organized to provide coverage of the subscribers. The area where the communication with the base station can be established is called a cell. Cells are usually assumed to be hexagonal regions with the base station being in the middle, and larger regions are assumed to be broken up into a hexagonal lattice of cells as seen from Figure \ref{fig:cell} \cite{andrea}.

Evolution of cellular networks has now reached its fourth generation. The cellular wireless generation (G) generally refers to a change in the fundamental nature of the service. The first generation mobile systems used analog transmission since they were primarily designed before the widespread use of digital communications.

In 1G, the forward channel (downlink) was used for transmissions from the base stations to mobile users, using frequencies between 869-894 MHz. Transmissions from mobile users to base station, on the other hand, occurred on the reverse channel (uplink), using frequencies between 824-849 MHz. Traffic was multiplexed onto an FDMA (frequency division multiple access) system. Frequency modulation (FM) technique was used by Advanced Mobile Phone System and Total Access Communication System for radio transmission \cite{v2libre3} \cite{v2libre5}.

Unlike 1G, second-generation (2G) systems used digital multiple access technology, such as TDMA (time division multiple access) and CDMA (code division multiple access). As a result, compared to 1G systems, 2G systems provided higher spectrum efficiency, better data services, and more advanced roaming.

3G does not consist of one standard; in fact, it is a family of standards that all work together. The organization called 3rd Generation Partnership Project (3GPP) has defined a mobile system that fulfils the IMT-2000 standard. 3G networks allow service providers to offer users more advanced services while attaining greater network capacity by means of improved spectral efficiency. 3G telecommunication networks support services that enable users to transfer data at a rate of at least 2Mbps.

The first successful trial of 4G was made in 2005. Users should be able to select the required wireless systems to use 4G services. In existent GSM systems, base stations periodically broadcast messages to mobile users for subscription. However, this procedure becomes more complex in 4G heterogeneous networks because of the differences in wireless technologies. Thus, terminal mobility is a requirement in 4G infrastructure to provide wireless services at anytime and anywhere. Terminal mobility allows mobile users to travel across boundaries of wireless networks. There are two main issues in terminal mobility: location management and handoff management. With location management, a mobile terminal is tracked and located by the system for a potential connection. Location management involves managing all the information about the roaming terminals, such as current and previous cells and authentication information. On the other hand, handoff management preserves the continuing communications when the terminal roams \cite{v2libre}.

The Third Generation Partnership Project (3GPP) had specified the basis of the future Long Term Evolution (LTE) Advanced, the 3GPP candidate for 4G, standards \cite{v2libre10}. The target values of peak spectrum efficiency for LTE Advanced systems were set to 30 bps/Hz for downlink and 15 bps/Hz for uplink channels. Improved multiple-input multiple-output (MIMO) channel transmission techniques and extensive coordination between multiple cell sites called coordinated multipoint (CoMP) transmission/reception were accepted as the key techniques for LTE, together with the multiple access schemes \cite{v2libre11}.

\subsubsection{Heterogeneous Networks}

\begin{figure}[thbp]
\begin{center}
\subfloat[2G/3G Typical Cellular Deployment \cite{figs}]{\includegraphics[width=0.5\textwidth]{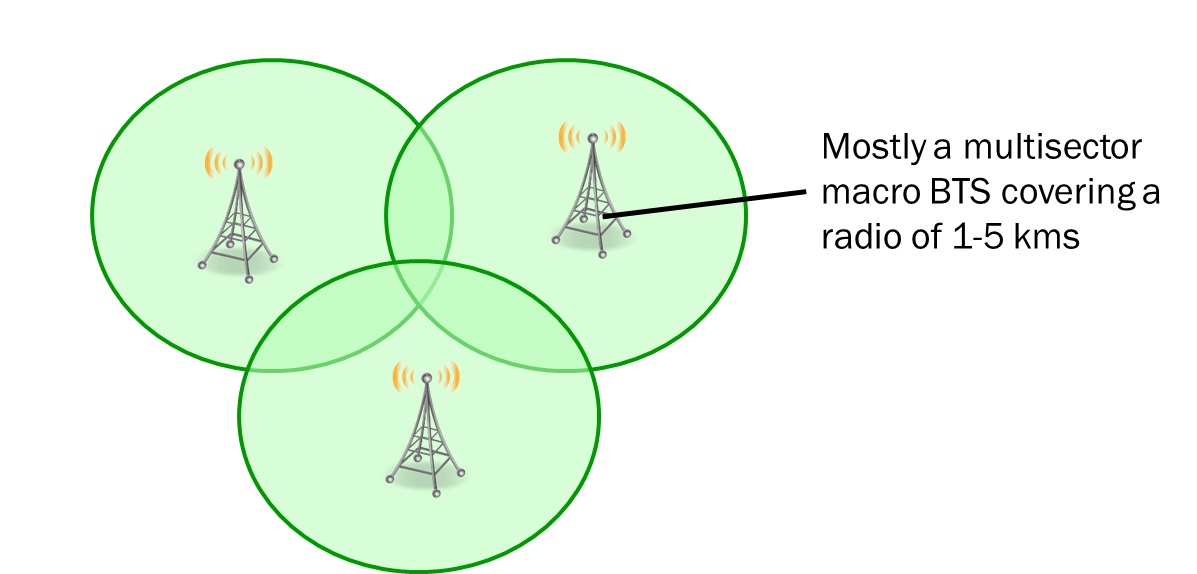}\label{fig:subfig1}}
\subfloat[4G HetNet Deployment \cite{figs}]{\includegraphics[width=0.5\textwidth]{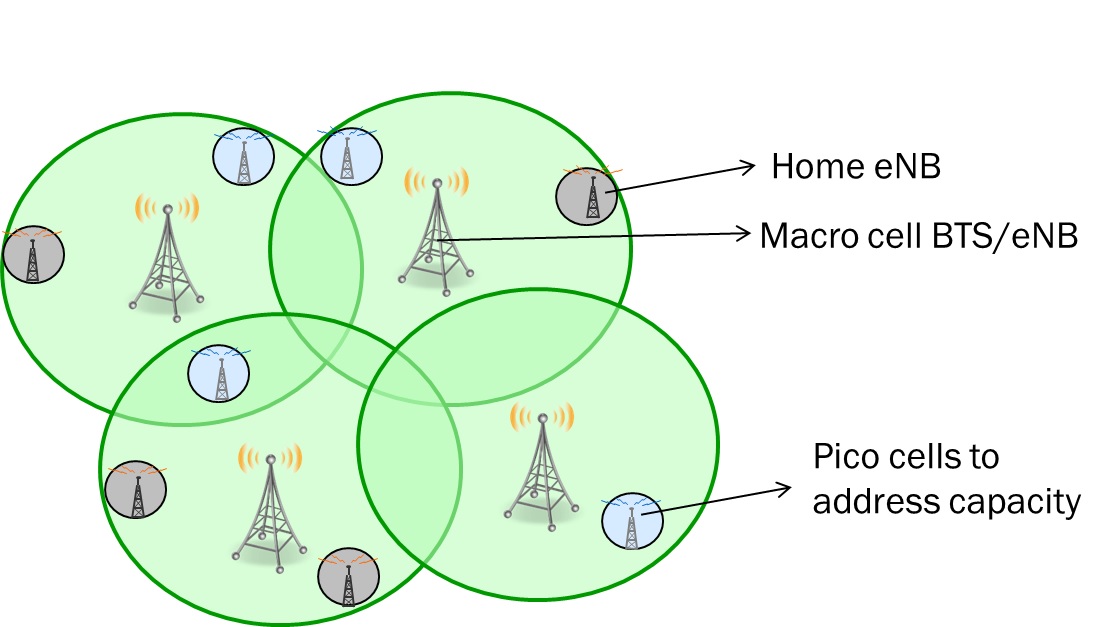}\label{fig:subfig2}}
\caption{Evolution of Cell Deployments}
\label{fig:4g}
\end{center}
\end{figure}

With LTE Advanced, a new term Heterogeneous Network (HetNet) was introduced and they have gained significant attention for their ability to optimize the system performance, especially for uneven user and traffic distribution. LTE networks were first based on homogeneous networks consisting of macro base stations that provided basic coverage. With the involvement of pico and femto base stations (eNBs), networks achieved significant improved overall capacity and cell-edge performance. In heterogeneous networks, layers of low power pico or femto eNBs (evolved node B) that are deployed in a less planned manner are on top of the layer of planned high power macro eNBs \cite{whitepaper}. Cellular systems in urban areas now generally use small cells for street level transmissions at much lower power \cite{andrea}. Typical 2G/3G and 4G HetNet deployments can be seen from Figure \ref{fig:4g} \subref{fig:subfig1} and \subref{fig:subfig2}, respectively.

However, the evolution of heterogeneous networks resulted in more complicated network designs. Since, mobile users pass through small cells more quickly than a macro cell; handoffs must be processed more quickly. Moreover, location management becomes harder, as there are more cells within a given area where a user can be located \cite{andrea}. Energy management is also an important issue, since activating pico nodes all the time may be energy inefficient. Thus, they should have some kind of sleep strategy.

\subsubsection{Wireless Communication Properties}

In wireless systems, output power of the transmitter is different from the power received by the receiver. This is mainly due to path loss, shadowing and multipath fading.
Path loss is usually expressed in dB. In its simplest form, the path loss can be calculated as
\begin{equation}
\begin{split}
\label{eq:pathloss}
L = 10 n \log_{10}d + C
\end{split}
\end{equation}
where $ L $ is the path loss in decibels, $ n $ is the path loss exponent, $ d $ is the distance between the transmitter and the receiver, usually measured in meters, and $ C $ is a constant which accounts for system losses.

Experiments reported by Egli in 1957 showed that, for paths longer than a few hundred meters, the received power fluctuates with a log-normal distribution about the area- mean power. This fluctuation is due to shadowing and the probability density function (pdf) of the local-mean power is thus of the form shown as
\begin{equation}
\label{eq:egli}
\begin{split}
f_{\bar{p}_{\log}}(\bar{p}_{\log}) = \frac{1}{\sqrt{2 \pi \sigma_{s}}} \exp [- \frac{1}{2 \sigma_{s}^{2}} {(\bar{p}_{\log})^{2}} ]
\end{split}
\end{equation}
where, $ \sigma_{s} $ is the logarithmic standard deviation of the shadowing, expressed in natural units.

Also, in wireless networks, the signal offered to the receiver contains not only a direct line-of- sight radio wave, but also a large number of reflected radio waves. The phases of the reflected waves are altered; thus, these reflected waves interfere with the direct wave, which causes significant degradation of the performance of the network known as multipath fading. Although channel fading is experienced as an unpredictable, stochastic phenomenon, powerful models such as Rician and Rayleigh Fading have been developed in order to accurately predict system performance \cite{harvard}.

\subsubsection{Spectral Efficiency and Energy Efficiency}

An important tool to analyse a network is quality of service (QoS). QoS is the overall performance of a network, particularly the performance experienced by the users of the network. To quantitatively measure quality of service, several related aspects of the network service are often considered, such as error rates, throughput, transmission delay and jitter. Quality of service can also be regarded as the ability to provide different priority to different applications, users, or data flows, or to guarantee a certain level of performance to a data flow. For example, a required bit rate, delay, jitter, packet dropping probability and/or bit error rate may be guaranteed \cite{wiki1}. In our case, some important QoS parameters may be bit rate and bits per joule.

To calculate the bit rate that a mobile user gets, we can use Shannon's capacity formula as 
\begin{equation}
\label{eq:shannon}
\begin{split}
C_{i} = w_{i} \log_{2}(1 + SNR_{i})
\end{split}
\end{equation}
where, $ w_{i} $ is the bandwidth of that user and $ SNR_{i} $ is the Signal to Noise Ratio of the user.

SNR that a user gets can be calculated by 
\begin{equation}
\label{eq:snr}
\begin{split}
SNR_{i} = \dfrac{P_{t}^{BS}  PL_{i}(d_{i})  G_{BS}  G_{MU}  P_{shadow}}{k  T  w_{i}}
\end{split}
\end{equation}
where $ P_{t}^{BS} $ is the transmit power of the base station, $ PL_{i}(d_{i}) $ is the path loss the user experiences, $ G_{BS} $ and $ G_{MU} $  are antenna gains of base station and mobile user, $ P_{shadow} $ is the shadowing with log-normal distribution, $ T $ is the temperature in Kelvin and $ k $ is the Boltzmann constant. $ SNR $ value in the receiver should be above a certain threshold for the symbols to be accurately interpreted.

Using these values and total consumed power, we can achieve an important parameter bits per joule, which is used as a metric for energy efficiency (EE). Bits per joule is simply the ratio of the capacity to the rate of energy expenditure \cite{3969}. It can be seen as a special case of the capacity per unit cost \cite{39692}. To calculate EE, we can use 
\begin{equation}
\label{eq:bpj}
\begin{split}
EE = \frac{\sum_{i} C_{i}}{P_{tot}} \, \, bits/J
\end{split}
\end{equation}
where the summation in the numerator is the total capacity of the network and $ P_{tot} $ is the total consumed power.

\section{System Model}
\label{system}
\subsection{Network Configurations}

LTE networks were first based on homogeneous networks consisting of macro base stations that provided basic coverage. That network configuration, consisting of macro cells only, is called MoNet.

\begin{figure}[thb]
\begin{center}
 \centerline{%
 \resizebox{0.5\textwidth}{!}{\includegraphics{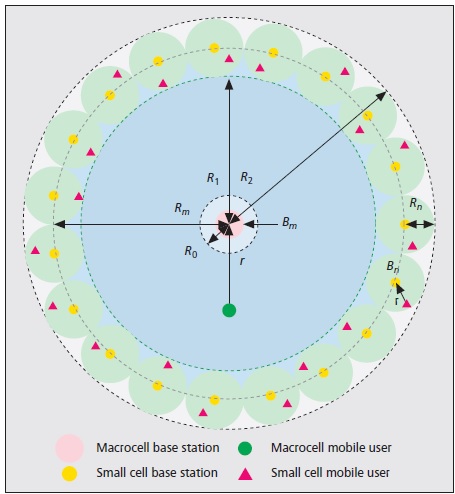}}%
 }
\caption{COE Configuration \cite{greenhet2}}

\label{fig:coe}
\end{center}
\end{figure}

Now that heterogeneous networks are seen as a promising way to meet the increasing demand for mobile broadband traffic in cellular networks, new pico nodes complement the macro nodes to provide higher capacity in areas with higher user density \cite{grc1}. There are two widely used topologies for heterogeneous networks, namely Cell on Edge (COE) and Uniformly Distributed Cells (UDC). In COE, small cells are located on the edge of the reference macro cell as in Figure \ref{fig:coe}. The main aim of the COE configuration is to provide capacity and coverage to cell edge mobile users by reducing the distance between the transmitter (mobile user in uplink) and the receiver (BS in uplink)\cite{greenhet2}. This way, since the mobile users that experience largest path loss are allowed to transmit to nearest small cell, capacities of the cell edge users increase. That leads to the increase of the overall capacity. However, since the layers of low-power pico nodes are less well planned or even entirely uncoordinated, COE is not a very realistic topology. 

\begin{figure}[thb]
\begin{center}
 \centerline{%
 \resizebox{0.45\textwidth}{!}{\includegraphics{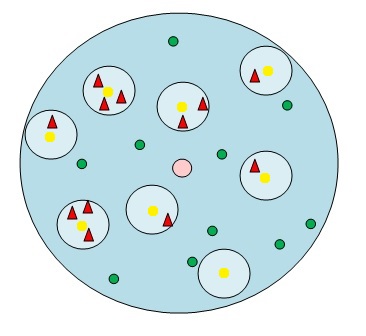}}%
 }
\caption{UDC Configuration. Yellow circles are small base stations, red triangles are the users served by the small cells, and the green circles are the users served by the macro base station.}

\label{fig:udc}
\end{center}
\end{figure}

Alternative approach is UDC, where the small cells are uniformly distributed across the macro cells as in Figure \ref{fig:udc} \cite{greenhet2}. UDC may not improve the system's capacity as much as COE, but it suits real life scenarios better.

\subsection{Base Station Power Models}

Recent surveys on the energy consumption of cellular networks, including base stations (BSs), mobile terminals and the core network, show that around 80\% of the energy required to run a cellular network is consumed at base stations \cite{enb5}. The energy efficiency frameworks therefore focus on the base station power consumption.

The base station power model maps the RF output power radiated at the antenna elements, $ P_{out} $, to the total supply power of a base station site, $ P_{in} $. In an LTE downlink, the eNB load is proportional to the amount of utilized resources, comprising both data and control signals. More generally the base station load also depends on power control settings, in terms of the transmitted spectral power density.

\begin{figure}[thb]
\begin{center}
 \centerline{%
 \resizebox{0.3\textwidth}{!}{\includegraphics{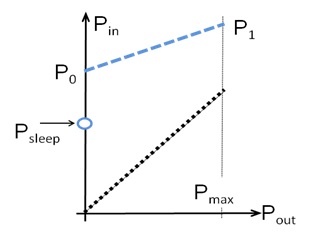}}%
 }
\caption{Load Dependent Power Model for a Typical LTE eNB \cite{enb}}

\label{fig:enb}
\end{center}
\end{figure}

Figure \ref{fig:enb} shows the principle characteristics of the proposed simplified estimate of an eNB power model for LTE. The detailed study of base station components in EARTH \cite{enb6} has shown that there is large offset power consumption at zero load. As indicated in Figure \ref{fig:enb}, the relations between relative RF output power $ P_{out} $ and eNB power consumption $ P_{in} $ are approximated by the function 
\begin{equation}
\label{eq:powercons}
\begin{split}
P_{in} = 
\begin{cases} 
      N_{trx}(P_{0} + \Delta_{p} P_{out}) & 0 < P_{out}\leq P_{max} \\
      N_{trx}\cdot P_{sleep} & P_{out} = 0 \\
\end{cases}
\end{split}
\end{equation}
where, $ P_{0} $  is the power consumption at the minimum non-zero output power, $ P_{max} $ is the maximum RF output power per component carrier and $ P_{out} $  is a fraction of it, and $ \Delta_{p} $ is the slope of the load dependent power consumption.

The parameters of the linear power model for the base stations of interest are obtained by least squares curve fitting of the detailed model \cite{enb6} and are listed in Table \ref{tab:bspow}. The parameters are based on an LTE system with 2x10 MHz bandwidth and 2x2 MIMO configuration \cite{enb}.

\begin{table}[ht]
\begin{center}
\caption{Power Model Parameters for Different Base Station Types \cite{earth}}
\label{tab:bspow}
\begin{tabular}{|c|c|c|c|c|c|}\hline
\textbf{BS type} & $ \mathbf{N_{TRX}} $& $ \mathbf{P_{max}[W]} $& $ \mathbf{P_{0}[W]} $& $ \mathbf{\Delta_{p}} $& $ \mathbf{P_{sleep}[W]} $ \\
\hline
Macro & 6 & 20 & 130 & 4.7 & 75  \\
\hline
Micro & 2 & 6.3 & 56 & 2.6 & 39  \\
\hline
Pico & 2 & 0.13 & 6.8 & 4.0 & 4.3 \\
\hline 
Femto & 2 & 0.05 & 4.8 & 8.0 & 2.9 \\
\hline 
\end{tabular}
\end{center}
\end{table}

\subsection{User Movements}

A suitable model for defining user movements is the Brownian Motion Model. Brownian motion is the random motion of particles suspended in a fluid (a liquid or a gas) resulting from their collision with the quick atoms or molecules in the gas or liquid \cite{brown}. The term "Brownian motion" can also refer to the mathematical model used to describe such random movements. The mathematical model of Brownian motion has numerous real-world applications, like stock market fluctuations.

An elementary example of a 1D Brownian motion is the random walk on the integer number line, $\mathbb{Z}$ , which starts at 0 and at each step moves +1 or −1 with equal probability. This walk can be illustrated as follows. A marker is placed at zero on the number line and a fair coin is flipped. If it lands on heads, the marker is moved one unit to the right. If it lands on tails, the marker is moved one unit to the left. After five flips, the marker could now be on 1, −1, 3, −3, 5, or −5. With five flips, three heads and two tails, in any order, will land on 1. There are 10 ways of landing on 1 (by flipping three heads and two tails), 10 ways of landing on −1 (by flipping three tails and two heads), 5 ways of landing on 3 (by flipping four heads and one tail), 5 ways of landing on −3 (by flipping four tails and one head), 1 way of landing on 5 (by flipping five heads), and 1 way of landing on −5 (by flipping five tails).

For the 2D illustration, we can imagine a drunkard walking randomly in an idealized city. The city is effectively infinite and arranged in a square grid, and at every intersection, the drunkard chooses one of the four possible routes (including the one he came from) with equal probability. Formally, this is a random walk on the set of all points in the plane with integer coordinates. The trajectory of this random walk is the collection of sites he visited \cite{random}.

For a map that that is not arranged in a grid topology, 2D Brownian motion can be modelled as follows: the users randomly select a destination coordinate inside the map and a speed (between 0 and pre-determined maximum). At each time slot, their positions are updated as $ X = X_{prev} + V_{x} $ and $ Y = Y_{prev} + V_{y} $. When they reach their destinations, they start the process again.

\subsection{User Distributions}

The users can be distributed within the network in two ways: uniform and non-uniform (hotspot). In uniform, user equipments (UEs) are randomly and uniformly distributed in the geographic coverage area of macro cells. On the other hand, in hotspot, a fraction of the total UEs are randomly placed within the coverage area low power cells and the remaining UEs are randomly and uniformly distributed within the macro cells \cite{qual}.

Hotspot is a more realistic approach than purely uniform distribution since mobile users are not uniformly distributed in real life. Instead, the number of users is denser in areas such as schools, shopping malls and hospitals. Some hotspots may periodically turn on and off, depending on the time of the day. For instance, a hotspot may occur in a business centre between 8am - 5pm, whereas another may operate in a shopping mall between 10am - 10pm. The small cells serving a hotspot may require more bandwidth than a regular small cell, in order to maintain the required QoS.

In this section, we have provided the related background information. In the next section, a scheduling problem that aims to maximize EE will be introduced and an activity management algorithm will be proposed.

\section{Problem Definition}
\label{chp:proposed}
Heterogeneous networks are seen as a promising way to meet the increasing demand for mobile broadband traffic in next generation cellular networks. The new pico-eNBs complement the macro-eNBs to provide higher capacity in areas with higher user density \cite{grc1}. That is because, even though they allocate the user same bandwidth that macro-eNB allocates, since their coverage area is smaller, their signals face lower path loss. Thus, the $ SNR $s that their users achieve, and bit rate in return, are larger. However, pico-eNB power model is as Equation (\ref{eq:powercons}). Therefore, with every eNB turning on, we encounter an offset power $ N_{trx} \cdot P_{0} $, independent from the number of users being served, and that offset power is comparable to, or even larger than, the user dependent power.

Thus, activating pico nodes all the time, for small number of users, is energy inefficient. Then, they should have some kind of sleep strategy. The pico-eNBs should be turned on and off in a way to maximize the overall EE, which is measured by $ bits/J $. Also, there should be a sufficient number of users in a pico cell for adequate improvement of the total capacity. For these purposes, we defined a hotspot model and turn on and off thresholds for pico-eNBs.

In short, the problem can be defined as determining the activation and deactivation thresholds, $ T_{activate} $ and $ T_{deactivate} $, in a way to maximize the EE, where EE is defined in Equation (\ref{eq:bpj}).

\subsection{5G Heterogeneous Networks}

5G infrastructure have important properties that allow HetNets to operate flawlessly. In this section, some of these properties are explained and fairness metric is discussed.

\subsubsection{Dual Connectivity}

The term "dual connectivity" is used to refer to operation where a given UE consumes radio resources provided by at least two different network points connected with non-ideal backhaul \cite{3gp}. That way, it is possible for a user to switch from macro cell to pico cell almost instantly, without interrupting the current transmissions. Thus, when a pico-eNB is turned off or when a UE enters the coverage area of a pico-eNB, it can instantly begin to use the resources of the new cell.

\subsubsection{Almost Blank Subframes}

Transmissions from macro-eNBs inflicting high interference onto pico-eNBs users are periodically muted (stopped) during entire subframes, this way the pico-eNB users that are suffering from a high level of interference from the aggressor macro-eNB have a chance to be served. However this muting is not complete as certain control signals are still transmitted which are:
\begin{itemize}
\item Common reference symbols (CRS) 
\item Primary and secondary synchronization signals (PSS and SSS)
\item Physical broadcast channel (PBCH)
\item SIB-113 and paging with their associated PDCCH.
\end{itemize}

These control channels have to be transmitted even in the muted subframes to avoid radio link failure or for reasons of backwards compatibility, so muted subframes should be avoided in subframes where PSS, SSS, SIB-1 and paging are transmitted or in other words subframes \#0, \#1, \#5 and \#9. Since these muted subframes are not totally blank they are called Almost Blank Subframes (ABS). The basic idea is to have some subframes during which the macro-eNB is not allowed to transmit data allowing the range extension pico-eNB users, who were suffering from interference from the macro-eNB transmission, to transmit with better conditions \cite{abs}.

\subsubsection{Fairness}

Among the fairness metrics such as max-min fairness, proportional fairness and $ \alpha $-fairness, we used "equal bandwidth share" that makes the users fair in terms of allocated bandwidth. The total allocated bandwidth of a macro cell is assumed to be $ W $. All pico and macro eNBs are assumed to be transmitting in different frequencies as it eliminates co-channel interference and provides better performance \cite{greenhet6}, so some of that bandwidth is given to pico-eNBs. The total number of users is assumed to be $ N $ and since they are fair in terms of allocated bandwidth, each user gets a bandwidth of $ W/N $. Then, the bandwidth allocated to a cell is directly proportional with the number of users within that cell.

\subsection{Activity Management Algorithm}

In the first network model, transmission energies of eNBs have been determined only according to path loss and path loss has been determined according to the Free Space Propagation Model, where the parameters are obtained from \cite{greenhet2}. The received signal power at a mobile user, from the desired eNB can be calculated by 
\begin{equation}
\begin{split}
P^{rx}(r) =  P^{tx} \dfrac{K}{r^{\alpha}(1 + r/g)^{\beta}}
\end{split}
\end{equation}
If we reverse this equation, we can achieve the desired transmit power at the base station as
\begin{equation}
\begin{split}
P^{tx}(r) = \min(P_{\max}, P_{0}\dfrac{r^{\alpha}(1 + r/g)^{\beta}}{K})
\end{split}
\end{equation}
where, $ r $ is the distance between user and base station, $ \alpha $ and $ \beta $ are basic and additional path loss exponents, $ g $ is the breakpoint distance, $ K $ is the path loss constant, $ P_{0} $ is the arbitrary cell-specific parameter that corresponds to the target signal-to-interference-plus-noise ratio, $ P_{max} $ is the maximum transmit power and $ P_{rx} $ and $ P_{tx} $ are received and transmit powers, respectively \cite{greenhet2}.

There have been no fixed energy parts that pico and macro-eNBs consumed. All stations have always been on, but of course, when they have not communicated with a user they have not been consuming energy, since the energy consumption due to node circuitry was ignored. In that case, there has been no cost of turning on and off the pico-eNBs. Also, there has been no downsides of serving under-utilized cells.

\begin{figure}[thb]
\begin{center}
 \centerline{%
 \resizebox{0.4\textwidth}{!}{\includegraphics{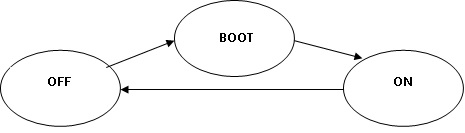}}%
 }
\caption{State Diagram of pico-eNBs}

\label{fig:states}
\end{center}
\end{figure}

However, this is not a realistic model since the base stations should also have some load independent energy. Also, they cannot start serving the users immediately after turning on. Therefore, in our more realistic network, the state diagram of pico-eNBs can be modelled as in Figure \ref{fig:states} and the energy consumption is as Equation (\ref{eq:powercons}), where there are both load dependent and load independent parts. So, for a pico-eNB to turn on, it should first enter boot mode. In boot mode, a node cannot communicate with users, it only consumes some fixed energy, representing the energy consumed by the circuitry. Boot mode lasts for some slot time. In that time, users communicate with macro-eNB, as pico-eNB is not yet ready for transmission. Since macro-eNB never goes into a sleep mode, no boot mode is required for it. 

\begin{figure}[thb]
\begin{center}
 \centerline{%
 \resizebox{0.7\textwidth}{!}{\includegraphics{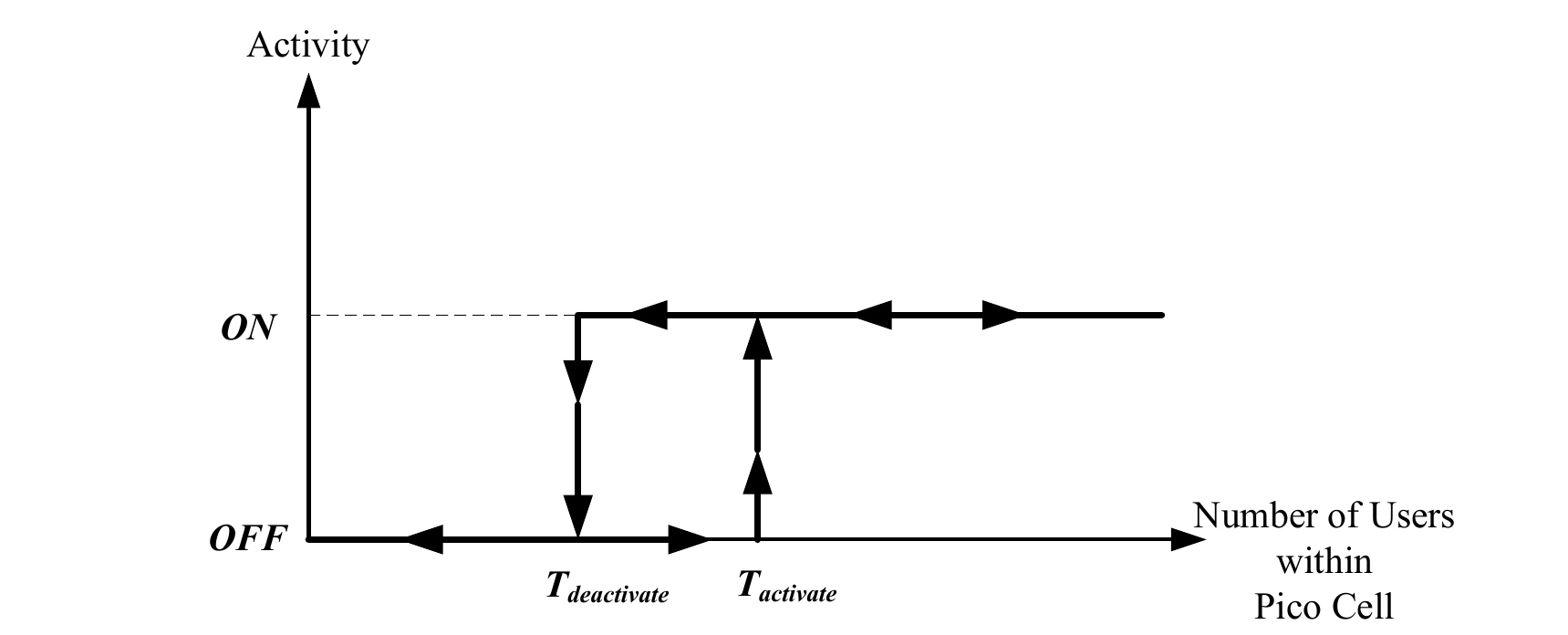}}%
 }
\caption{Hysteresis Shape of Activity Model of pico-eNBs}
\label{fig:hyst}
\end{center}
\end{figure}

The main objective is to maximize the EE and thus, optimally decide the thresholds for a pico-eNB to turn on and turn off. They are meant to be optimum in the sense of energy usage and network capacity. We would like to turn the pico-eNBs on, in order to increase the capacity. However, we would also like to turn off under-utilized eNBs to decrease the consumed energy, as the eNBs consume high fixed powers. In addition, these two thresholds should be different. Otherwise, the node may oscillate between two modes, as the node cannot go from sleep to active mode instantly, and that causes delay and wasted energy. Thus, number of users versus consumed power diagram will be shaped as hysteresis as in Figure \ref{fig:hyst}.

However, in the case that users do not enter and leave pico cells instantaneously, only one threshold is of interest; since in that case, the pico-eNB do not turn on and off quite often and probability of oscillation is very low. In return, the location of the second threshold would not change the performance of the system. However, even in that case, the two thresholds may require being different if both the conditions involve the equality. Because when they are equal in slot n, if no user enters or leaves a pico cell with number of users being $ T_{activate} $, it would oscillate; since, $ T_{activate} = T_{deactivate} $ and eNB cannot decide whether to turn on or off. As an alternative, we can define only one threshold and arrange the conditions such that only one condition involves the equality and the other would require the number of users to be strictly less than or larger than that threshold. That way, when the thresholds are equal in slot n, even if no user enters or leaves a pico cell with number of users being $ T_{acttivate} $, it still would not oscillate, as the condition would require the number of users within the cell to be strictly less than $ T_{activate} $, for the eNB to turn off.

\subsubsection{Hotspot Model}

EE of the users of pico cells may be sufficiently large. However, number of users in pico cells may be very small, since when the users are distributed randomly, their probability of being inside a pico cell is quite little. Thus, the overall effect of the pico cells would not be very much. Then, we understand that to see the benefit of this heterogeneous network more, we should increase the number of users in pico cells. For that purpose, instead of distributing the users randomly, we should introduce a hotspot concept. Then, a fraction of the total UEs would be randomly placed within 50 meters around the pico-eNB and the remaining UEs would be randomly and uniformly distributed within the macro cell \cite{qual}.

As an example to the hotspot model, one can think of business centres and shopping malls. Thus, major number of users enter and leave the pico cells together, at opening and closing hours. That's why, when we use hotspot model, one threshold can be sufficient both for turning on and off the eNBs. There may be other users entering and leaving pico cells instantaneously, but they would not cause the eNB to switch between on and off modes. 

In this section, we have introduced the problem of maximizing EE and described the proposed activity management algorithm and the employed hotspot model in detail. In the next section simulation environment including network topology, user distributions, mobility, activity, energy and channel models will be explained. Then, the simulation results will be provided and explained in detail.

\section{Simulation Results}
\label{chp:simulation}

In this section, simulation environment and parameters used are introduced. Then, simulation results are provided and obtained results are discussed in detail.

\subsection{Network Topology}

\begin{figure}[thb]
\begin{center}
 \centerline{%
 \resizebox{0.7\textwidth}{!}{\includegraphics{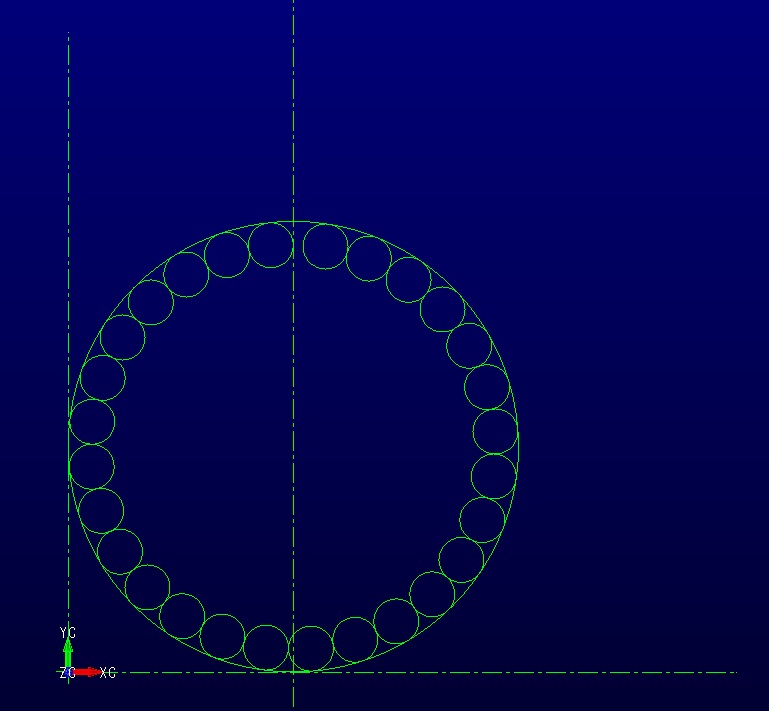}}%
 }
\caption{COE Configuration Used in Simulations}

\label{fig:ediz}
\end{center}
\end{figure}

As mentioned in section \ref{chp:proposed}, there are 3 network configurations that we consider: COE, UDC and MoNet. There are 28 pico cells both in UDC and COE with coverage radius being 50m, whereas the coverage radius of macro cell is 500m. In real applications, the pico eNBs are planned to be placed in only concentration areas for optimum performance; thus, their number per macro cell can be larger or smaller than this value. The locations of pico eNBs in UDC configuration is random such that no two cells intersect. On the other hand, their locations in COE  are specially designed so that they cover almost all areas at the edge of macro cell. The COE configuration that is used in simulations can be seen from Figure \ref{fig:ediz}.

In the simulations, we have considered only one macro cell and it is placed so as to be tangential to the axes. Then, the pico cells are positioned in a manner that they would be tangential to both the macro cell and each other. After that, the locations of macro and pico eNBs are determined and used in simulations.

\subsubsection{User Distributions}
\subsubsection{Uniform}

In the uniform user distribution, the users are placed to their initial locations randomly. There are 1000 users. For any pico cell, if the distance between UE and eNB is less than 50m, then the user is assumed to be within that pico cell. When the distance between the UE and eNB exceeds 50m, the user is immediately served by macro eNB. No slot is wasted during this process, thanks to dual connectivity. In this case, since the users are distributed in a uniform manner, the maximum number of users that a pico cell can get is around 20. This number is a little low to see the benefits of the heterogeneous network.

In order to make a fair comparison between network configurations, the users are distributed and moved the same in all three cases. In other words, UDC, COE and MoNet configurations can be visualised as transparent layers and users are located below them. Thus, all configurations see the same users in same locations at all times.

\subsubsection{Hotspot}
\label{chp:hotspot}
In hotspot scenario, there are $ n $ hotspot users and $ 1000-n $ random users that act like in uniform case. All hotspot users are assigned a specific pico cell and work time during their creations. Then, initially, all users are again distributed randomly. When their assigned work time come, each user picks a random location from their allocated pico cell and starts moving towards it. In this case, when we have taken the number of hotspot users as 500, the maximum number of users that a pico cell can get is around 50, which is a sufficient number to see the benefits of heterogeneous network as shown by simulations.

However, since the cells are located in different places in UDC and COE, user distributions cannot be the same in hotspot scenario for both UDC, COE and MoNet. Thus, to make a fair comparison, we have taken the UDC and COE hotspot user distributions for MoNet, as two different scenarios. In other words, in these cases, MoNet can be visualized as a transparent layer on top of UDC and COE user distributions. Also, we named these two scenarios as MoNet with COE configuration and MoNet with UDC configuration. Naturally, MoNet with UDC configuration performs better than MoNet with COE configuration, since in the latter, macro eNB needs to serve many users at the edge of its coverage area.

\subsection{Mobility Model}

All users move according to the Brownian Motion Model, with different parameters. Uniform users choose a speed between 0 and 20 randomly and at each slot, they update their positions as $ X = X_{prev} + V_{x} $ and $ Y = Y_{prev} + V_{y} $. On the other hand, hotspot users choose a random speed between 10 and 20 when they start moving towards their assigned pico cell on their allocated time slot. When they reach their cell, their velocity drops to a random number between 0 and 2 and they choose a different destination within that cell. When their working time elapse, they choose a completely random new destination that can be within or outside the cell. This is a valid assumption since we associate UEs with real people who goes to work at specific times and then remain inside their workplaces for some time, where they move with much lower speeds.

Two pseudocodes that explain the initialization and user movements are shown in Algorithm \ref{alg1} and Algorithm \ref{alg2}, respectively. Hotspot users run both algorithms as they are. On the other hand, random users do not choose a pico cell in the initialization. Also, in the user movement algorithm, since random users do not have a work time, they skip the conditions related to work time and only check if they reached their destinations.

\begin{algorithm}
\caption{Initialization}
\label{alg1}
\begin{algorithmic}[1]
\State $ \textit{MyPico} \gets \text{random pico cell} $
\State $ \textit{x-coordinate} \gets \text{random within macro cell} $
\State $ \textit{y-coordinate} \gets \text{random within macro cell} $
\State $ \textit{x-destination} \gets \text{random within macro cell} $
\State $ \textit{y-destination} \gets \text{random within macro cell} $
\State $ \textit{MySpeed} \gets \text{random between 10m/slot and 20m/slot} $
\State $ \textit{MySpeedX} \gets \text{projection of the speed on x-coordinate} $
\State $ \textit{MySpeedY} \gets \text{projection of the speed on y-coordinate} $
\end{algorithmic}
\end{algorithm}

\begin{algorithm}
\caption{User Movement}
\label{alg2}
\begin{algorithmic}[2]
\IF {work time comes}
\State $ \textit{x-destination} \gets \text{random within} MyPico $
\State $ \textit{y-destination} \gets \text{random within} MyPico $
\State $ \textit{MySpeed} \gets \text{random between 10m/slot and 20m/slot} $
\State $ \textit{MySpeedX} \gets \text{projection of the speed on x-coordinate} $
\State $ \textit{MySpeedY} \gets \text{projection of the speed on y-coordinate} $
\ENDIF
\IF {work time ends}
\State $ \textit{x-destination} \gets \text{random within macro cell} $
\State $ \textit{y-destination} \gets \text{random within macro cell} $
\State $ \textit{MySpeed} \gets \text{random between 10m/slot and 20m/slot} $
\State $ \textit{MySpeedX} \gets \text{projection of the speed on x-coordinate} $
\State $ \textit{MySpeedY} \gets \text{projection of the speed on y-coordinate} $
\ENDIF
\State $ \textit{x-coordinate} \gets \textit{x-coordinate} + \textit{MySpeedX} $
\State $ \textit{y-coordinate} \gets \textit{y-coordinate} + \textit{MySpeedY} $
\IF {destination reached}
\IF {within work time}
\State $ \textit{x-destination} \gets \text{random within} MyPico $
\State $ \textit{y-destination} \gets \text{random within} MyPico $
\State $ \textit{MySpeed} \gets \text{random between 0m/slot and 2m/slot} $
\State $ \textit{MySpeedX} \gets \text{projection of the speed on x-coordinate} $
\State $ \textit{MySpeedY} \gets \text{projection of the speed on y-coordinate} $
\ELSE 
\State $ \textit{x-destination} \gets \text{random within macro cell} $
\State $ \textit{y-destination} \gets \text{random within macro cell} $
\State $ \textit{MySpeed} \gets \text{random between 10m/slot and 20m/slot} $
\State $ \textit{MySpeedX} \gets \text{projection of the speed on x-coordinate} $
\State $ \textit{MySpeedY} \gets \text{projection of the speed on y-coordinate} $
\ENDIF
\ENDIF
\end{algorithmic}
\end{algorithm}

\subsection{Activity Model}

All users make the decision of being active or idle in the next time slot. Uniform users are active with probability 0.4 and hotspot users are active with probability 0.8, when they are within a pico cell. The idle users transmit no data to the eNB and in return, when turning on an eNB, we only consider active users inside the cell. When the user is active, we assume that it continuously transmits packets throughout that slot.

In the case of MoNet, the users have the same activity model with the corresponding user distribution. In other words, in the case of MoNet with UDC user distribution, the users that are inside pico cells in UDC configuration, are again active with probability 0.8. We have made this assumption since the activity model of a user is independent from the location of pico cells. In fact, in reality, the cells are placed in locations where the users are active most. In other words, a man uses internet more at work, independent from whether there is a hotspot at work or not.

\subsection{Channel Model}

Channel parameters such as path loss formulas and shadow fading standard deviations are obtained from \cite{3gp}, in addition to parameters such as cell radius, transmit powers, total bandwidth and antenna gains. These parameters are summarized in Table \ref{tab:parame}.

\begin{table}[ht]
\begin{center}
\caption{Power Model Parameters for Different Base Station Types \cite{3gp}}
\label{tab:parame}
\begin{tabular}{|l|p{7cm}|}\hline
\textbf{Parameters} & \textbf{Settings/Assumptions} \\
\hline \hline
Cell radius & Macro cell: 500m \newline Small cell: 50m \\
\hline
Transmit power & Macro eNB: 46dBm \newline Small cell: 30dBm \\
\hline
Bandwidth & $ 2\times10 $MHz @ 2GHz and 3.5GHz \\
\hline 
Antenna configuration & $ 2\times2 $ MIMO with rank adaptation and interference rejection \\
\hline 
Antenna gain & Macro: 14dBi \newline Small cell: 5dBi \\
\hline 
Path loss & Macro cell: $ 140.7+36.7\log10(R[km]) $ \newline Small cell: $ 128.1+37.6\log10(R[km]) $ \\
\hline 
Shadow fading & Macro cell: lognormal, std=8dB \newline Small cell: lognormal, std=10dB \\
\hline 
\end{tabular}
\end{center}
\end{table}

\subsection{Energy Model}
\label{chp:ene}

In the simulations, one day corresponds to 1000 slots. There are 3 different work times for hotspot users that begin in 0\textsuperscript{th}, 42\textsuperscript{nd} and 83\textsuperscript{rd} slots. The users remain at work for 375 slots, that corresponds to 9 hours. When the number of users within a pico cell exceeds the turn on threshold, the pico eNB first enters a boot mode that lasts for 1 slot and then turns on. The energy consumption parameters that are used in simulations are decided with the help of \cite{earth} and \cite{3gp}. Table \ref{tab:bspow} summarizes the parameters in \cite{earth} and Table \ref{tab:paramet} shows the parameters of \cite{3gp}. We took the values that they both agree on, which means $ P_{sleep} = 8.6 W $, $ P_{0} = 13.6 W $, $ P_{max} = 0.25 W $ and $ \Delta_{p} = 4 $. The formula that is used to calculate the consumed power is expressed in Equation (\ref{eq:powercons}). For the boot mode, we assumed that the eNB consumes $ P_{sleep} $ and it cannot serve any users for that slot.

\begin{table}[ht]
\begin{center}
\caption{Energy Consumption Parameters for Various Base Station Types \cite{3gp}}
\label{tab:paramet}
\begin{tabular}{|l|l|l|l|l|l|l|}\hline
\textbf{BS type} & $ N_{sec} $ & \multicolumn{2}{|c|}{$ P_{max} $ \newline [W] [dBm]} & $ P_{0} $ [W] & $ \Delta_{p} $ & $ P_{sleep} $ [W] \\
\hline \hline
Macro & 3 & 40.0 & 46 & 260 & 4.75 & 150 \\
\hline
\multirow{4}{*}{RRH} & 3 & 40 & 46 & 168.0 & 2.8 & 112.0 \\
 & 1 & 5.0 & 37 & 103.0 & 6.5 & 69.0 \\
 & 1 & 1.0 & 30 & 96.2 & 1.5 & 62.0 \\
 & 1 & 0.25 & 24 & 13.6 & 4.0 & 8.6 \\ \hline
\multirow{3}{*}{Pico} & 1 & 5.0 & 37 & 103.0 & 6.5 & 69 \\
 & 1 & 1.0 & 30 & 96.2 & 1.5 & 62.0 \\
 & 1 & 0.25 & 24 & 13.6 & 4.0 & 8.6 \\
\hline 
Femto & 1 & 0.1 & 20 & 9.6 & 8.0 & 5.8 \\
\hline 
\end{tabular}
\end{center}
\end{table}

Even though $ P_{sleep} = 8.6 W $ in Table \ref{tab:paramet}, we have also simulated the case where $ P_{sleep} $ gets smaller values. That is because, when the eNBs do not serve users for some time, they have a chance to enter a deeper sleep mode and in that case $ P_{sleep} = 8.6 W $ is a high value. However, there is not much numerical information regarding different sleep mode energy consumption parameters in the literature; thus, we have compared several cases with various $ P_{sleep} $ values.

There are several suggestions to sleep modes such as; small cell controlled sleep mode, core network controlled sleep mode and UE controlled sleep mode. In small cell controlled sleep mode, by using the existence of macro cell coverage, the small cell hardware can be enriched with a low-power sniffer capability that allows the detection of an active call from a UE. Then, the small cell can afford to disable its pilot transmissions and the associated radio processing when no active calls are being made by the UE in its coverage area. In core network controlled sleep mode, different from the suggestion above, this case does not require the low-power sniffer in the small cell to detect active UEs. Alternatively, the transition of small cell from sleep to active state is controlled by the core network via the backhaul using a wake-up control message \cite{sleep}.

Our regular sleep mode is small cell controlled sleep mode and the deeper sleep mode can be assumed as core network controlled sleep mode. That way, in the deeper sleep mode, since the pico eNB do not require sniffing constantly, it consumes less power.

\subsection{Results}

\begin{figure}[thb]
\begin{center}
 \centerline{%
 \resizebox{0.7\textwidth}{!}{\includegraphics{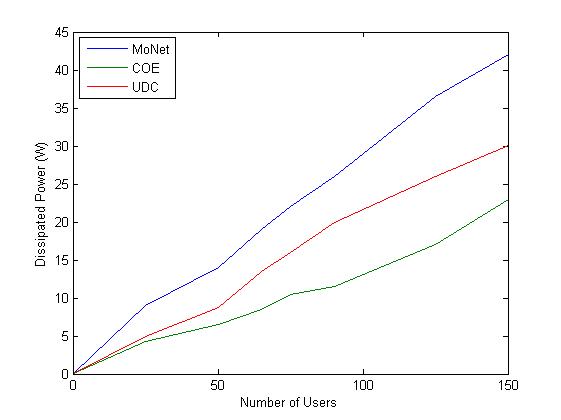}}%
 }
\caption{Dissipated Power vs. Number of Users}
\label{fig:first}
\end{center}
\end{figure}

In our first simulation, we have compared the total energy consumptions of MoNet, COE and UDC with respect to number of users and achieved Figure \ref{fig:first}. 

In this network model, transmission energies of eNBs are determined only according to path loss and path loss is determined according to Free Space Model, where the parameters are obtained from \cite{greenhet2} and explained in Table \ref{tab:param}. Here, we have assumed that all users are uniformly distributed.

\begin{table}[ht]
\begin{center}
\caption{Initial Simulation Parameters \cite{greenhet2}}
\label{tab:param}
\begin{tabular}{|c|c|c|}\hline
\textbf{Simulation Parameter} & \textbf{Small cell} & \textbf{Macro cell}\\
\hline \hline
Transmit power ($ P_{max} $) & 1W & 1W \\
\hline
Cell radius ($ R $) & 50m & 500m  \\
\hline
Path loss exponent ($ \alpha $) & 1.8 & 2 \\
\hline 
Additional path loss exponent ($ \beta $) & 1.8 & 2 \\
\hline 
BS antenna height ($ h_{BS} $) & 12.5m & 25m \\
\hline 
Mobile antenna height ($ h_{MU} $) & 2m & 2m \\
\hline 
Target SNR ($ P_{0} $) & $ 0.8\mu W $ & $ 0.8\mu W $ \\
\hline 
Breakpoint distance ($ g $) & 300m & 600m \\
\hline 
System bandwidth ($ W $) & \multicolumn{2}{|c|}{20MHz} \\
\hline 
Path loss constant ($ K $) & \multicolumn{2}{|c|}{1} \\
\hline 
\end{tabular}
\end{center}
\end{table}

As seen from Figure \ref{fig:first}, dissipated power increases with increasing number of users, since the total transmission energy is the sum of energies required to communicate with each user. The graph starts at point $ (0, 0) $, as there is no load-independent power consumption of macro and pico eNBs. Here, we observe that MoNet performs worst, since the path loss that users face in that topology is much larger than other cases and transmission energies are determined according to path loss. Also, we notice that COE performs better than UDC; since in COE, the users that would face largest path loss are served by the pico eNB, that is located at a smaller distance. That way, the path loss is significantly reduced.

\begin{figure}[thb]
\begin{center}
 \centerline{%
 \resizebox{0.7\textwidth}{!}{\includegraphics{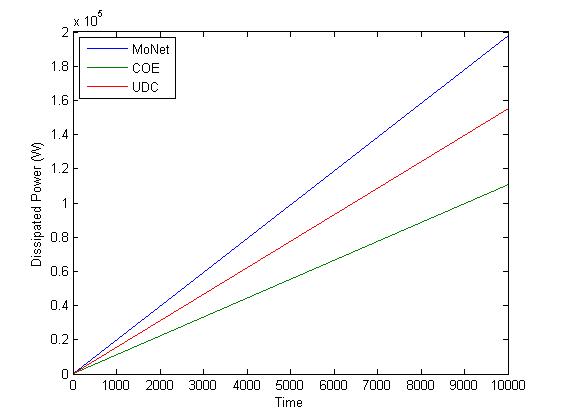}}%
 }
\caption{Dissipated Power vs. Time for 150 Stationary Users}
\label{fig:stat}
\end{center}
\end{figure}

\begin{figure}[thb]
\begin{center}
 \centerline{%
 \resizebox{0.7\textwidth}{!}{\includegraphics{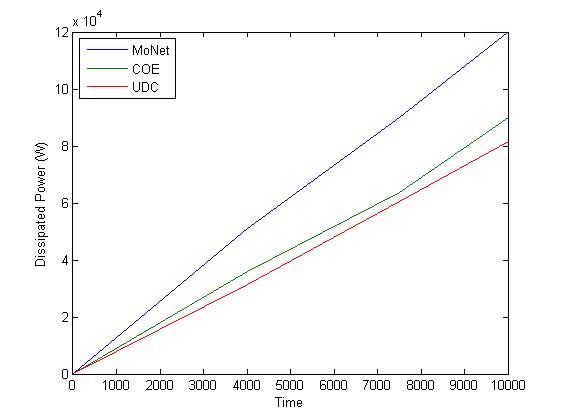}}%
 }
\caption{Dissipated Power vs Time for 150 Users with a Maximum Speed of 5 m/slot}
\label{fig:5user}
\end{center}
\end{figure}

\begin{figure}[thb]
\begin{center}
 \centerline{%
 \resizebox{0.7\textwidth}{!}{\includegraphics{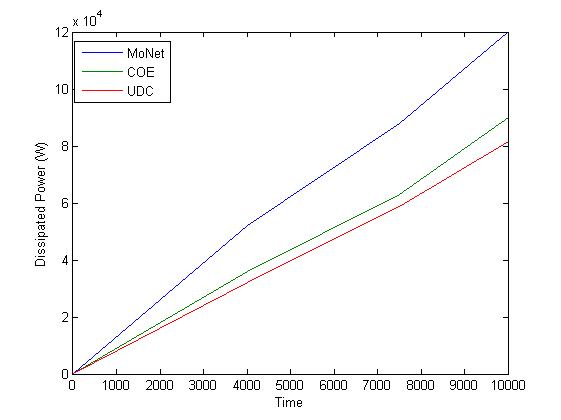}}%
 }
\caption{Dissipated Power vs Time for 150 Users with a Maximum Speed of 10 m/slot}
\label{fig:10user}
\end{center}
\end{figure}

\begin{figure}[H]
\begin{center}
 \centerline{%
 \resizebox{0.7\textwidth}{!}{\includegraphics{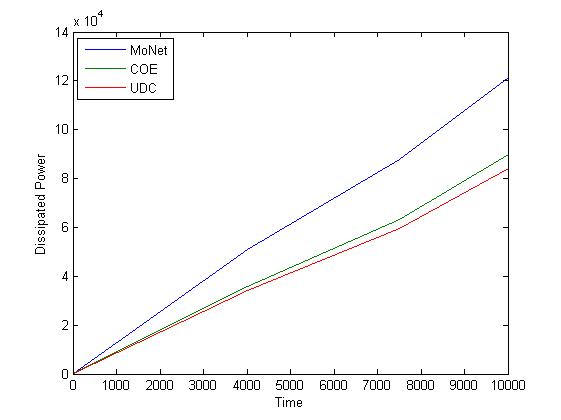}}%
 }
\caption{Dissipated Power vs Time for 150 Users with a Maximum Speed of 20 m/slot}
\label{fig:20user}
\end{center}
\end{figure}

After that, the users are made mobile and the system is switched to a slotted one. Then, the simulation is rerun for the users with different maximum speeds and the resulting graphs are shown in Figure \ref{fig:stat}, Figure \ref{fig:5user}, Figure \ref{fig:10user}, and Figure \ref{fig:20user}. Again, the users are distributed uniformly.

These graphs show the dissipated power with respect to time for 150 users. In the case of mobile users, UDC performed better than COE and favoured slightly more with higher speeds. That is because, in case of stationary users, the maximum energy gain is achieved when the users which are furthest from the macro eNB use pico eNBs; since the transmit energy is determined according to Free Space Model. However, in case of mobile users, since it is less likely for a user to travel on the edges of the simulation area than to travel near the centre, pico cells are less utilized in COE case. Thus, the energy gain for that case is less than UDC.

Then, we have changed the energy model to the model explained in Section \ref{chp:ene}. After that, dissipated powers of different base stations have been calculated using Equation \ref{eq:powercons} and found as
\begin{equation}
\label{eq:enerjisay1}
\begin{split}
P_{macro} = 780 + \dfrac{120 \times 4.75 \times n_{user}}{1000}
\end{split}
\end{equation}
\begin{equation}
\label{eq:enerjisay2}
\begin{split}
P_{pico} = 13.6 + \dfrac{0.25 \times 4 \times n_{user}}{50}
\end{split}
\end{equation}
where the maximum user capacities of macro and pico eNBs are assumed to be 1000 and 50, respectively. We need to assign some values to these parameters, since in Equation \ref{eq:powercons}, $ P_{out} $ is a fraction of $ P_{max} $, which depends on the ratio number of users served over number of maximum users. 

When we have changed the energy model, as the fixed energies of eNBs are larger than or comparable to their transmission energies, MoNet has become more efficient than COE and UDC in terms of total consumed power. Thus, to see the benefit of the heterogeneous network, we have investigated the parameter $ bits/J $. 

In the next simulation, we have assumed that the nodes in sleep mode consume no energy. Again, the users are distributed uniformly and they are assumed to be active with probability 1. Here, all the pico eNBs are in sleep mode initially. In this simulation, we consider 100 realizations of one time slot. If the number of users they serve is larger than the threshold in that slot, the nodes switch to active mode. $ bits/J $, capacity and total dissipated power values for three different turn-on thresholds are inspected in Table \ref{tab:sim}.

\begin{table}[ht]
\begin{center}
\caption{$ bits/J $, Total Capacity and Dissipated Energy Values for Various Thresholds}
\label{tab:sim}
\begin{tabular}{ |l|l|l|l|l| }
\hline
&& $ \mathbf{Bits/J} $ & \textbf{Capacity (bits)} & \textbf{Energy (J)} \\ \hline
\multirow{3}{*}{\textbf{Threshold = 0}} & MoNet & 340577 & $ 4.5977\times10^{8} $ & 1350 \\
 & COE & 325110 & $ 5.0493\times10^{8} $ & 1553.12\\
 & UDC & 314784 & $ 4.9512\times10^{8} $ & 1572.91\\ \hline
\multirow{3}{*}{\textbf{Threshold = 8}} & MoNet & 339764 & $ 4.5868\times10^{8} $ & 1350 \\
 & COE & 335670 & $ 4.9476\times10^{8} $ & 1473.96\\
 & UDC & 322315 & $ 4.8694\times10^{8} $ & 1510.77\\ \hline
\multirow{2}{*}{\textbf{Threshold = 13}} & MoNet & 339948 & $ 4.5892\times10^{8} $ & 1350\\
 & COE & 342915 & $ 4.7430\times10^{8} $ & 1383.15\\
 & UDC & 342026 & $ 4.6691\times10^{8} $ & 1365.15\\
\hline
\end{tabular}
\end{center}
\end{table}

As seen from Table \ref{tab:sim}, as we increase the threshold, total capacity decreases; since the number of users being served by pico eNBs lessens. However, when we turn off underutilized cells, we begin to outperform MoNet, in terms of $ bits/J $. That is because, even though we turn off some cells, we still serve some users within active pico cells and that provides a gain due to path loss. We also get rid of the zero-load powers of underutilized cells.

\begin{figure}[thb]
\begin{center}
 \centerline{%
 \resizebox{0.7\textwidth}{!}{\includegraphics{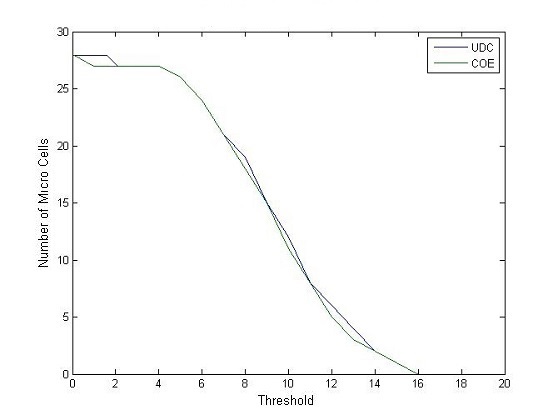}}%
 }
\caption{Number of Active Pico eNBs vs. Threshold}
\label{fig:openmicros}
\end{center}
\end{figure}

Then, we have rerun the simulation to find the optimum threshold value. Again, we consider 100 realizations of one time slot. The number of active pico cells for different thresholds are as seen from Figure \ref{fig:openmicros}. When the threshold exceeds 18, there are no more active pico cells. This is expected, since the users are uniformly distributed.

\begin{figure}[thb]
\begin{center}
 \centerline{%
 \resizebox{0.7\textwidth}{!}{\includegraphics{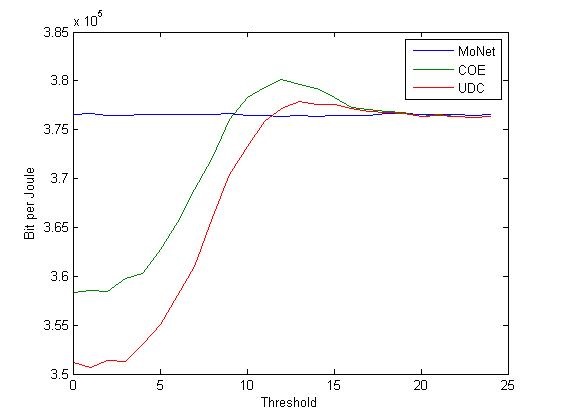}}%
 }
\caption{$ bits/J $ vs. Threshold for MoNet, UDC and COE with $ P_{sleep}  = 0 W $}
\label{fig:eksik}
\end{center}
\end{figure}

The change in the $ bits/J $ with respect to threshold can be seen from the Figure \ref{fig:eksik}. As seen here, the optimum turn on threshold is 12 for COE and 13 for UDC, since they have the largest $ bits/J $ ratios. Also, COE performs better than UDC, since here, we only see the effect of path loss and not mobility, as we only consider 1 time slot.

\begin{figure}[thb]
\begin{center}
 \centerline{%
 \resizebox{0.7\textwidth}{!}{\includegraphics{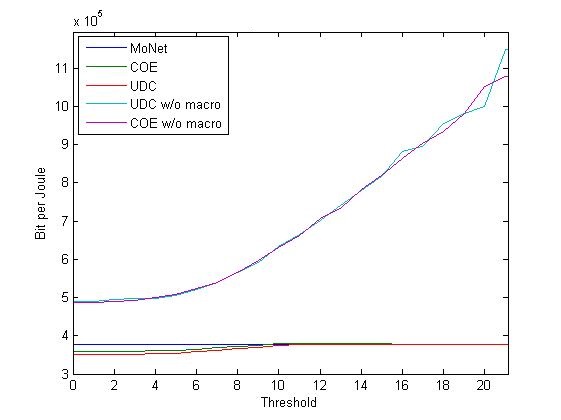}}%
 }
\caption{$ bits/J $ vs. Threshold for MoNet, UDC, COE, UDC w/o Macro and COE w/o Macro with $ P_{sleep}  = 0 W $}
\label{fig:sl0hs0}
\end{center}
\end{figure}

Then, to see the effect of pico cells more, on top of the $ bits/J $ graph of COE, UDC and MoNet, we have added COE and UDC without the energy and bit rate of macro cell. After that, the resulting graph became as in Figure \ref{fig:sl0hs0}.

As seen from the figure, $ bits/J $ of the users of pico cells are considerably large. However, the number of users in pico cells is very small. So, the overall effect of the pico cells is not sufficient. From that, we understand that to see the benefit of the heterogeneous network more, we should increase the number of users in pico cells. 

\begin{figure}[thb]
\begin{center}
 \centerline{%
 \resizebox{0.7\textwidth}{!}{\includegraphics{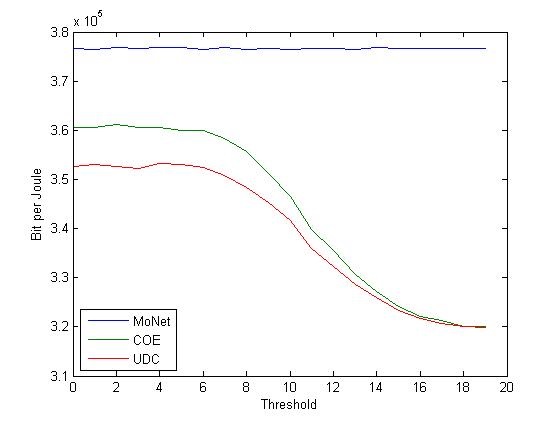}}%
 }
\caption{$ bits/J $ vs. Threshold for MoNet, UDC and COE with $ P_{sleep}  = 8.6 W $}
\label{fig:sleepli1}
\end{center}
\end{figure}

\begin{figure}[thb]
\begin{center}
 \centerline{%
 \resizebox{0.7\textwidth}{!}{\includegraphics{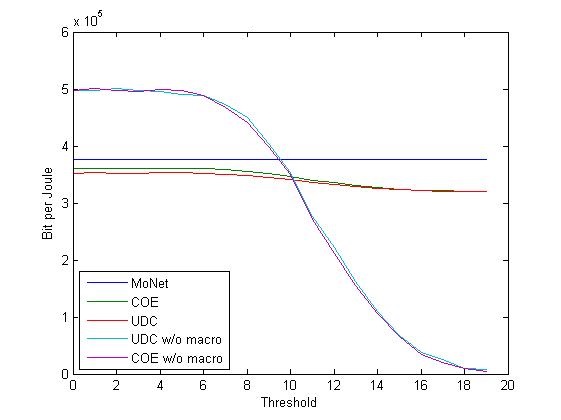}}%
 }
\caption{$ bits/J $ vs. Threshold for MoNet, UDC, COE, UDC w/o Macro and COE w/o Macro with $ P_{sleep}  = 8.6 W $}
\label{fig:sleepli2}
\end{center}
\end{figure}

However, these results are for $ P_{sleep} = 0 W $, as expressed before. When we take $ P_{sleep} = 8.6 W $, which is the value provided in \cite{enb6}, Figure \ref{fig:eksik} and \ref{fig:sl0hs0} transform into Figure \ref{fig:sleepli1} and \ref{fig:sleepli2}.

As seen from Figure \ref{fig:sleepli1} and \ref{fig:sleepli2}, since the sleep energies of the nodes are quite high, when we turn off underutilized nodes, their large sleep energies cause a significant decrease in overall $ bits/J $. Thus, we should either implement a sleep mode with a smaller sleep energy or increase the number of users within pico cells so that more pico cells will be turned on at any time and they will serve a suitable number of users. For that purpose, instead of distributing the users randomly, we should introduce the hotspot concept.

\begin{figure}[thb]
\begin{center}
 \centerline{%
 \resizebox{0.7\textwidth}{!}{\includegraphics{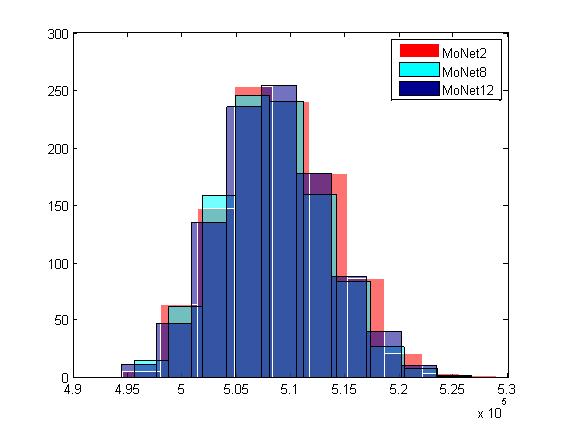}}%
 }
\caption{Histogram of Capacities of MoNet Users for Thresholds 2,8 and 12}
\label{fig:monetbits}
\end{center}
\end{figure}

\begin{figure}[thb]
\begin{center}
 \centerline{%
 \resizebox{0.7\textwidth}{!}{\includegraphics{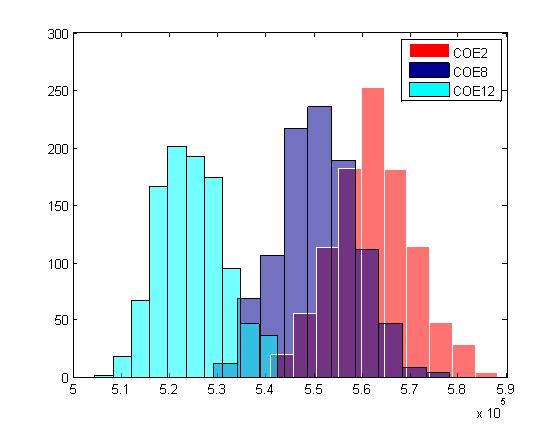}}%
 }
\caption{Histogram of Capacities of COE Users for Thresholds 2,8 and 12}
\label{fig:coebits}
\end{center}
\end{figure}

\begin{figure}[thb]
\begin{center}
 \centerline{%
 \resizebox{0.7\textwidth}{!}{\includegraphics{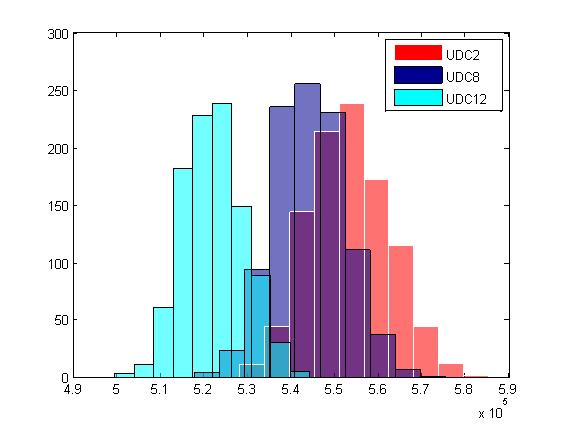}}%
 }
\caption{Histogram of Capacities of UDC Users for Thresholds 2,8 and 12}
\label{fig:udcbits}
\end{center}
\end{figure}

In addition, to see the fairness between users in terms of bit rates, the histogram of bit rates have been drawn for thresholds 2, 8 and 12. Once again, we consider 100 realizations of one time slot. The histograms of users in MoNet, COE and UDC are as as shown in Figures \ref{fig:monetbits}, \ref{fig:coebits} and \ref{fig:udcbits} respectively.

In MoNet, the change in the thresholds do not change the histogram of capacities, since there are no pico eNBs to turn on and off. But in UDC and COE, as the threshold increases, users get less capacities as expected, since fewer number of users are able to use pico cells, when threshold gets higher. For instance, when the threshold is 2 and most pico cells are active, some users in COE and UDC get a capacity around $ 5.9 \times 10^{5} $ b/s. Also, on the avarage, they get around $ 5.6 \times 10^{5} $ b/s. However, as the threshold increases and reaches 12, the highest capacity that the users get becomes $ 5.45 \times 10^{5} $ b/s and the avarage capacity becomes $ 5.25 \times 10^{5} $ b/s. These values are very close to MoNet case, since when the threshold is high, most pico eNBs are turned off.

\begin{figure}[thb]
\begin{center}
 \centerline{%
 \resizebox{0.7\textwidth}{!}{\includegraphics{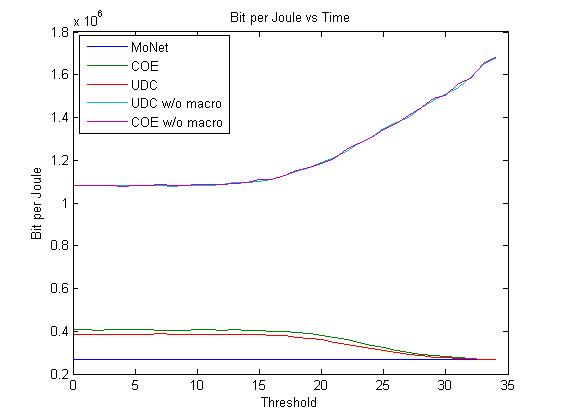}}%
 }
\caption{$ bits/J $ vs. Threshold for MoNet, UDC, COE, UDC w/o Macro and COE w/o Macro with $ P_{sleep}  = 0 W $ and Hotspot Users = 500}
\label{fig:sleep0}
\end{center}
\end{figure}

\begin{figure}[thb]
\begin{center}
 \centerline{%
 \resizebox{0.7\textwidth}{!}{\includegraphics{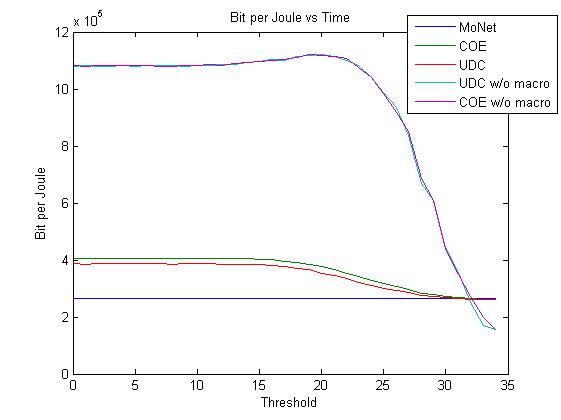}}%
 }
\caption{$ bits/J $ vs. Threshold for MoNet, UDC, COE, UDC w/o Macro and COE w/o Macro with $ P_{sleep}  = 2 W $ and Hotspot Users = 500}
\label{fig:sleep2}
\end{center}
\end{figure}

\begin{figure}[thb]
\begin{center}
 \centerline{%
 \resizebox{0.7\textwidth}{!}{\includegraphics{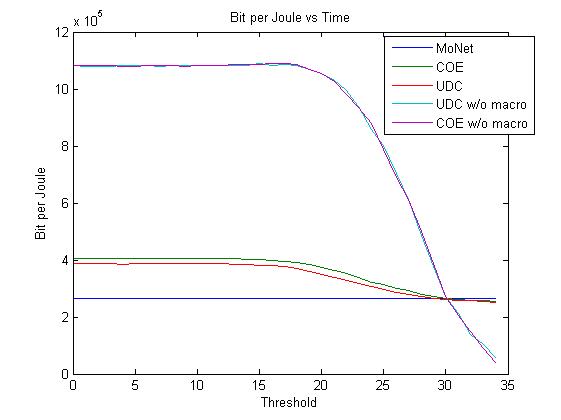}}%
 }
\caption{$ bits/J $ vs. Threshold for MoNet, UDC, COE, UDC w/o Macro and COE w/o Macro with $ P_{sleep}  = 4 W $ and Hotspot Users = 500}
\label{fig:sleep4}
\end{center}
\end{figure}

\begin{figure}[H]
\begin{center}
 \centerline{%
 \resizebox{0.7\textwidth}{!}{\includegraphics{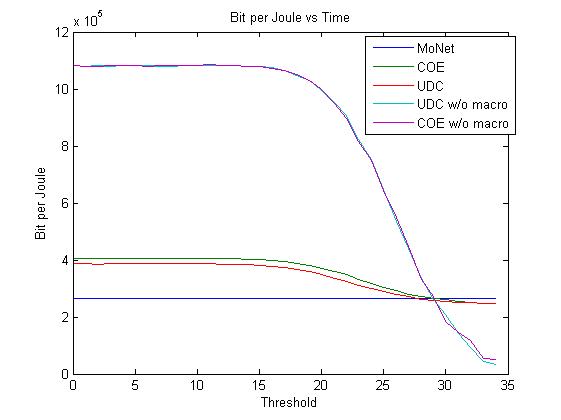}}%
 }
\caption{$ bits/J $ vs. Threshold for MoNet, UDC, COE, UDC w/o Macro and COE w/o Macro with $ P_{sleep}  = 6 W $ and Hotspot Users = 500}
\label{fig:sleep6}
\end{center}
\end{figure}

\begin{figure}[thb]
\begin{center}
 \centerline{%
 \resizebox{0.7\textwidth}{!}{\includegraphics{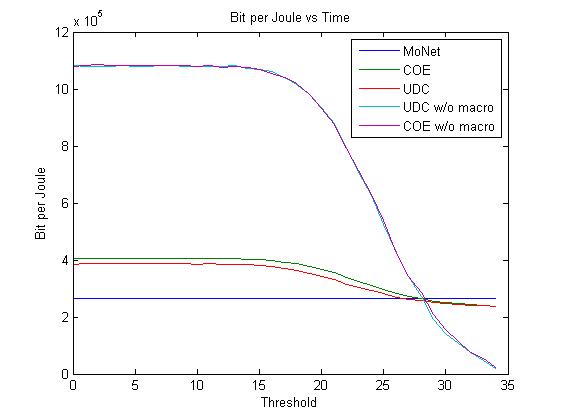}}%
 }
\caption{$ bits/J $ vs. Threshold for MoNet, UDC, COE, UDC w/o Macro and COE w/o Macro with $ P_{sleep}  = 8.6 W $ and Hotspot Users = 500}
\label{fig:sleep8}
\end{center}
\end{figure}

In the next simulations, we have introduced the hotspot model and compared the $ bits/J $ of MoNet, COE and UDC users, varying sleep energies and number of users in hotspots. While varying the sleep energies of eNBs, we have assumed the number of hotspot users to be 500 ;and while varying the number of hotspot users, we have considered two sleep energies: 0 and 8.6. In these simulations, again we have considered 100 realizations of one time slot. Figures \ref{fig:sleep0}, \ref{fig:sleep2}, \ref{fig:sleep4}, \ref{fig:sleep6} and \ref{fig:sleep8} show the performances of different topologies with respect to threshold for various sleep energies.

As we see from the figures, as the sleep energies of pico eNBs increase, $ bits/J $ of COE and UDC users start to drop below the $ bits/J $ of MoNet users, for higher thresholds. That is because, when the threshold is below 15, no pico eNB is in sleep mode; thus sleep energies of pico eNBs do not affect the performance. However, when the threshold exceeds 25, many of the pico cells begin to turn off and consume sleep energy. As a result, when most or all of them are turned off, the users begin to be served by macro cell, but pico eNBs consume additional sleep power. Then, the $ bits/J $ of COE and UDC users drop below the $ bits/J $ of MoNet users.

\begin{figure}[thb]
\begin{center}
 \centerline{%
 \resizebox{0.7\textwidth}{!}{\includegraphics{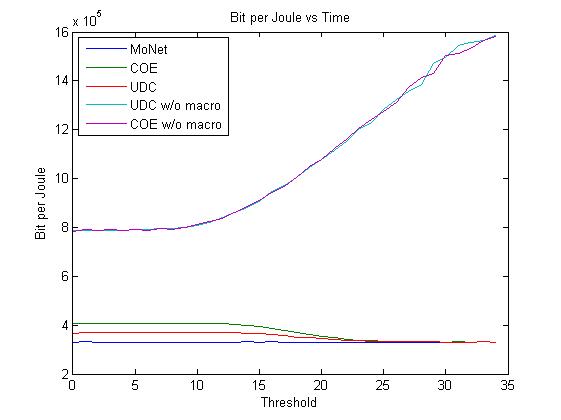}}%
 }
\caption{$ bits/J $ vs. Threshold for MoNet, UDC, COE, UDC w/o Macro and COE w/o Macro with $ P_{sleep}  = 0 W $ and Hotspot Users = 250}
\label{fig:sl0hs250}
\end{center}
\end{figure}

\begin{figure}[thb]
\begin{center}
 \centerline{%
 \resizebox{0.7\textwidth}{!}{\includegraphics{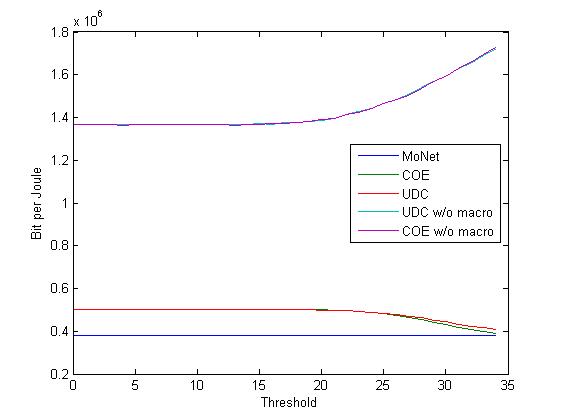}}%
 }
\caption{$ bits/J $ vs. Threshold for MoNet, UDC, COE, UDC w/o Macro and COE w/o Macro with $ P_{sleep}  = 0 W $ and Hotspot Users = 750}
\label{fig:sl0hs750}
\end{center}
\end{figure}

Then, we have inspected the effect of number of hotspot users on the network performance in terms of $ bits/J $. First, we have taken $ P_{sleep} = 0 W $, and varied the number of hotspot users from 0 to 750. We have already given the results of the cases where hotspot users = 0 and hotspot users = 500 in Figures \ref{fig:sl0hs0} and \ref{fig:sleep0}. The resulting graphs of the cases where hotspot users = 250 and hotspot users = 750 are shown in Figures \ref{fig:sl0hs250} and \ref{fig:sl0hs750}, respectively. 

\begin{figure}[thb]
\begin{center}
 \centerline{%
 \resizebox{0.7\textwidth}{!}{\includegraphics{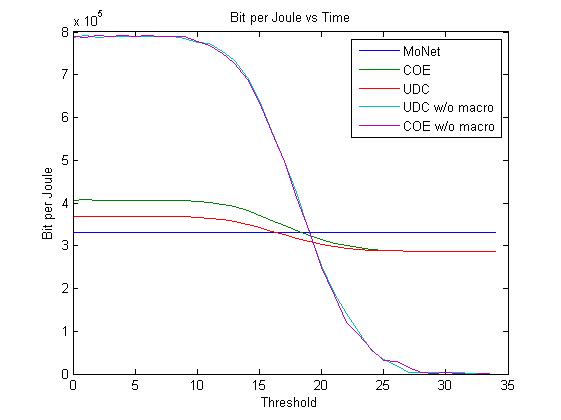}}%
 }
\caption{$ bits/J $ vs. Threshold for MoNet, UDC, COE, UDC w/o Macro and COE w/o Macro with $ P_{sleep}  = 8.6 W $ and Hotspot Users = 250}
\label{fig:sl8hs250}
\end{center}
\end{figure}

\begin{figure}[H]
\begin{center}
 \centerline{%
 \resizebox{0.7\textwidth}{!}{\includegraphics{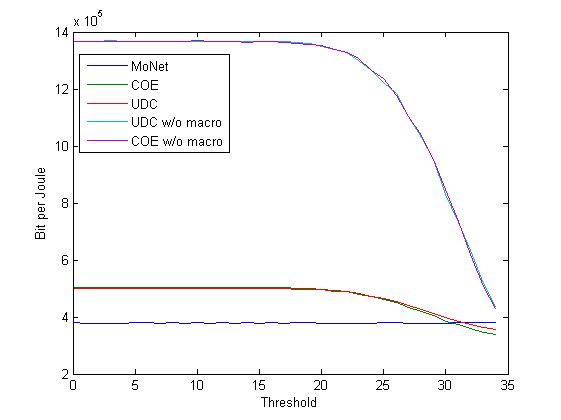}}%
 }
\caption{$ bits/J $ vs. Threshold for MoNet, UDC, COE, UDC w/o Macro and COE w/o Macro with $ P_{sleep}  = 8.6 W $ and Hotspot Users = 750}
\label{fig:sl8hs750}
\end{center}
\end{figure}

After that, we have taken $ P_{sleep} = 8.6 W $, and varied the number of hotspot users from 0 to 750. Again, we have given the results of the cases where hotspot users = 0 and hotspot users = 500 in Figures \ref{fig:sleepli1} and \ref{fig:sleep8}. The resulting graphs of the cases where hotspot users = 250 and hotspot users = 750 are shown in Figures \ref{fig:sl8hs250} and \ref{fig:sl8hs750}, respectively.

When we compare these figures, we observe that as the number of hotspot users increase, the difference between HetNet scenarios and MoNet increases. Thus, our aim should be maximizing the number of hotspot users and minimizing the sleep energies of pico eNBs. We may not change the dissipated power of the nodes in sleep mode; however, we may define a deeper sleep mode with less sleep energy and less features of nodes. Also, we can increase the number of hotspot users by placing the pico eNBs to the areas with higher user density.

We also observe that in the case with hotspot users, the turn on threshold becomes less meaningful. Because, when there are hotspot users, the more active the pico eNBs are, the more UDC and COE outperforms MoNet. As the pico eNBs begin the turn off, the performance of UDC and COE approaches to, or even becomes worse than, the performance of MoNet. Thus, in that case, it is better to turn on all the pico eNBs at all times.

\begin{figure}[thb]
\begin{center}
 \centerline{%
 \resizebox{0.7\textwidth}{!}{\includegraphics{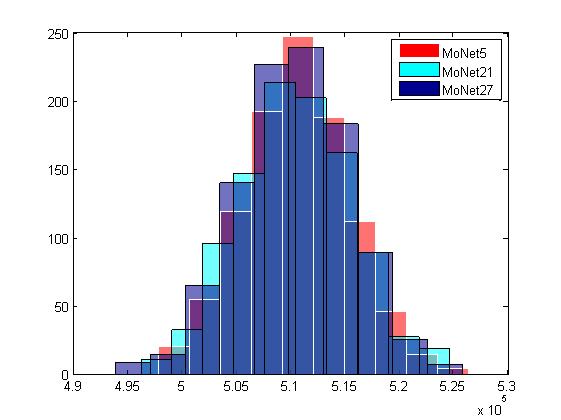}}%
 }
\caption{Histogram of Capacities of MoNet Users for Thresholds 5,21 and 27, and Hotspot Users = 500}
\label{fig:monethot}
\end{center}
\end{figure}

\begin{figure}[thb]
\begin{center}
 \centerline{%
 \resizebox{0.7\textwidth}{!}{\includegraphics{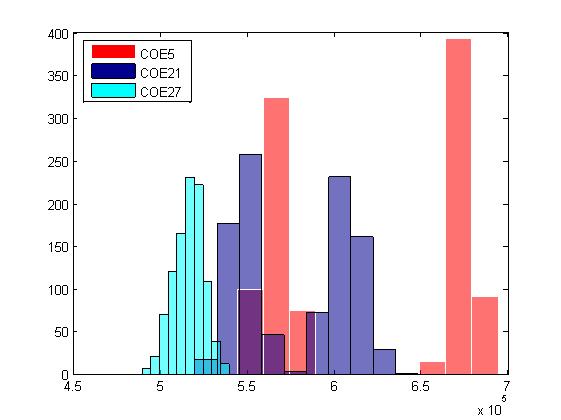}}%
 }
\caption{Histogram of Capacities of COE Users for Thresholds 5,21 and 27, and Hotspot Users = 500}
\label{fig:coehot}
\end{center}
\end{figure}

\begin{figure}[thb]
\begin{center}
 \centerline{%
 \resizebox{0.7\textwidth}{!}{\includegraphics{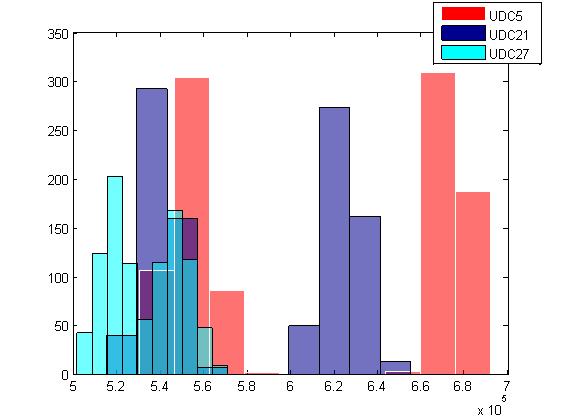}}%
 }
\caption{Histogram of Capacities of UDC Users for Thresholds 5,21 and 27, and Hotspot Users = 500}
\label{fig:udchot}
\end{center}
\end{figure}

In addition, to see the fairness between users in terms of bit rates, the histogram of bit rates have been drawn for thresholds 5, 21 and 27, and hotspot users = 500. These thresholds are higher than before, since in the case of hotspot users, the number of users within pico cells are higher. Thus, to see the effect of the threshold, we must select it higher. Once again, we consider 100 realizations of one time slot. The histograms of users in MoNet, COE and UDC are as as shown in Figure \ref{fig:monethot}, \ref{fig:coehot} and \ref{fig:udchot} respectively. Here, $ P_{sleep} = 8.6 W $, but this does not make a difference in the bit rates.

In MoNet, the change in the thresholds do not change the histogram of capacities, since there are no pico eNBs to turn on and off. But in UDC and COE, as the threshold increases, users get less capacities as expected, since fewer number of users are able to use pico cells, when threshold becomes higher. On the other hand, when the threshold is low, the users in UDC and COE achieve noticeably higher bit rates, compared to MoNet. Also, in Figures \ref{fig:coehot} and \ref{fig:udchot}, we observe that when the threshold is 5 or 21, the histogram is split. This is because, there is a difference between the bit rates of the users that are served by macro and pico eNBs, and when there are 500 hotspot users, the number of users served by pico eNBs are quite significant. Thus, the bit rates of the users that are served by pico eNBs are considerably higher than the bit rates of the users that are served by macro eNBs. However, when the threshold is 27, since the number of users served by pico eNBs is very small, the bit rates they get do not appear apart from the other users. This split histograms are the main difference from the histograms in Figures \ref{fig:monetbits}, \ref{fig:coebits} and \ref{fig:udcbits}; since in that case, number of pico cell users were much smaller than the hotspot case.

\begin{figure}[thb]
\begin{center}
 \centerline{%
 \resizebox{0.7\textwidth}{!}{\includegraphics{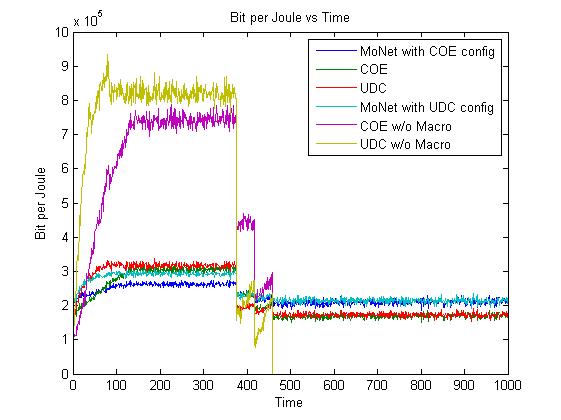}}%
 }
\caption{$ bits/J $ vs. Time for MoNet, UDC, COE, UDC w/o Macro and COE w/o Macro with $ P_{sleep}  = 0 W $ and Hotspot Users = 500}
\label{fig:all}
\end{center}
\end{figure}

\begin{figure}[thb]
\begin{center}
 \centerline{%
 \resizebox{0.7\textwidth}{!}{\includegraphics{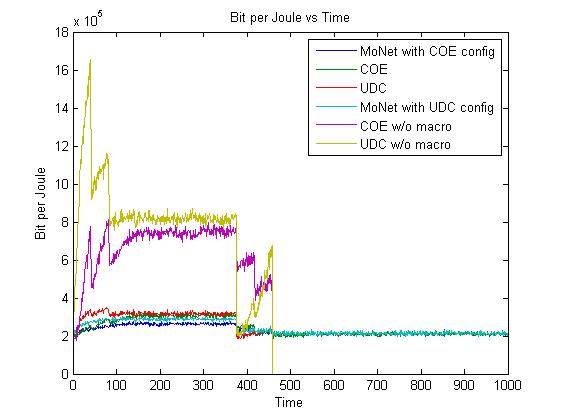}}%
 }
\caption{$ bits/J $ vs. Time for MoNet, UDC, COE, UDC w/o Macro and COE w/o Macro with $ P_{sleep}  = 8.6 W $ and Hotspot Users = 500}
\label{fig:all2}
\end{center}
\end{figure}

In the next simulation, we have compared the performances of the topologies with respect to time. We have run the simulation for 1000 slots, which corresponds to 1 day. As expressed before, users are mobile with a random speed between 10-20 m/slot. There are 500 hotspot and 500 randomly distributed users. Inside a pico cell, hotspot users are active with probability 0.8, and outside they are active with probability 0.4, as well as random users. The working times of hotspot users begin at slots 0,42 and 83; and then, they begin moving towards their assigned pico cells. The change in the $ bits/J $ with respect to time for $ P_{sleep} = 0 W $ and $ P_{sleep} = 8.6 W $ are shown in Figures \ref{fig:all} and \ref{fig:all2}, respectively. MoNet with COE configuration and MoNet with UDC configuration are as defined in \ref{chp:hotspot}.

In Figure \ref{fig:all}, pico cells have no sleep energy; thus, when they turn off, COE and UDC have the same $ bits/J $ with MoNet. However, in Figure \ref{fig:all2}, the sleep energies of the cells are assumed as 8.6. Thus, when the pico cells turn off, the $ bits/J $ of UDC and COE fall below MoNet. 

Since all users do not arrive their pico cells at the same time, the increase in the $ bits/J $ at the beginning is not sharp. On the other hand, during turn off, as the pico cells stop serving users instantly, $ bits/J $ drops right away. Also, the $ bits/J $ of COE reach its maximum value later than UDC. That is because the pico cells are located at the edges of the simulation area and it takes longer time to reach them.

Here, UDC has higher $ bits/J $, since both UDC and COE provide approximately the same $ bits/J $ to pico cell users, but random users have a higher probability to be served by UDC. Also, there is around \%20 improvement for both cases. To find the improvements, we have compared UDC with MoNet with UDC configuration and COE with MoNet with COE configuration.

\begin{figure}[thb]
\begin{center}
 \centerline{%
 \resizebox{0.7\textwidth}{!}{\includegraphics{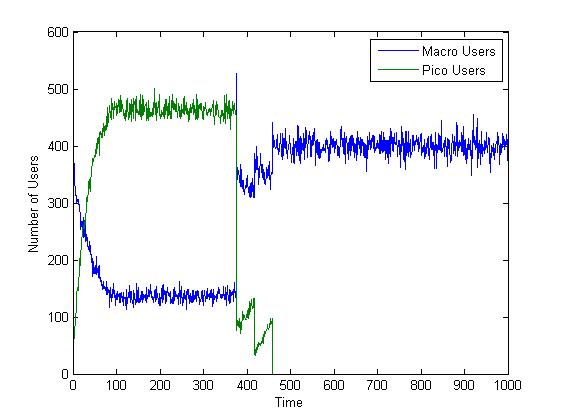}}%
 }
\caption{Number of Users Served by Pico and Macro Cells vs. Time in UDC}
\label{fig:sonudc}
\end{center}
\end{figure}

\begin{figure}[H]
\begin{center}
 \centerline{%
 \resizebox{0.7\textwidth}{!}{\includegraphics{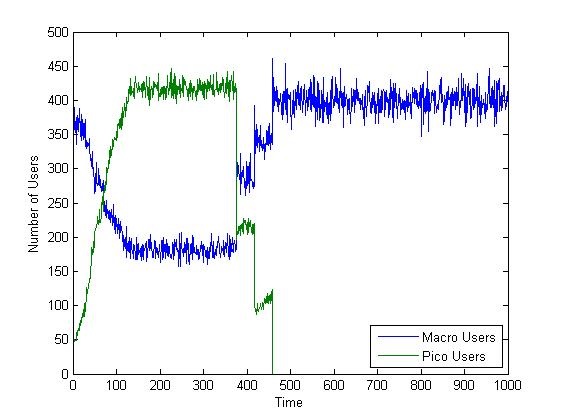}}%
 }
\caption{Number of Users Served by Pico and Macro Cells vs. Time in COE}
\label{fig:soncoe}
\end{center}
\end{figure}

Then, we have inspected the number of users served by pico and macro eNBs in UDC and COE. The resulting graphs are shown in Figures \ref{fig:sonudc} and \ref{fig:soncoe}. Here, hotspot users = random users = 500.

In Figure \ref{fig:sonudc}, we can see the number of active users with respect to time. Number of active macro users start from 400; since, when all the pico cells are turned off, all 1000 users are connected to macro eNB and they are active with probability 0.4. Around 100th slot, all assigned users are inside their pico cells and active with probability 0.8, thus, number of pico users should be at least 400. They are around 470, since some of the randomly distributed users are also within pico cells. Finally, around 470\textsuperscript{th} slot, pico cells are turned off and all 1000 users are connected to macro eNB again. As they are active with probability 0.4, number of active macro users become around 400.

In Figure \ref{fig:soncoe}, number of active macro users start from 400, too. On the contrary to the UDC, here the assigned users reach their pico cells around 150\textsuperscript{th} slot, since the pico cells are on the edges of the simulation area. Then, the number of active users in pico cells reach around 410. That is because, the number of random users being served by pico eNBs is very low, since they are located at the end of the coverage area of macro cell. When pico eNBs start to turn off, the number of users inside the pico cells do not drop as fast as UDC; since in the case of COE, when a user leaves a pico cell, it is very likely to enter another pico cell, as they are located side by side. And when the user enters the neighbour cell, that cell may not be turned off yet. Thus, the total number of users being served by pico eNBs do not drop as fast as in UDC. Finally, around 470\textsuperscript{th} slot, pico cells are turned off and all 1000 users are connected to macro eNB again. As they are active with probability 0.4, number of active macro users become around 400.

Then, in order to see difference between the one-threshold system and the two-threshold system, we have rerun the simulation for the sleep energies $ P_{sleep} = 0 $ and $ P_{sleep} = 8.6 $, and various thresholds. When there is only one threshold, we assume that the pico cell turns on when the number of users within the cell is larger than or equal to the threshold; and it turns off when the number of users gets strictly less than the threshold. When there are two thresholds, we assume that the pico cell turns on when the number of users within the cell is larger than or equal to the $ T_{activate} $; and it turns off when the number of users gets less than or equal to $ T_{deactivate} $.

\begin{figure}[thb]
\begin{center}
 \centerline{%
 \resizebox{0.7\textwidth}{!}{\includegraphics{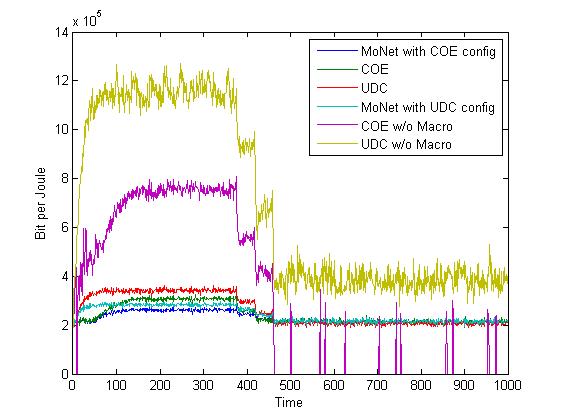}}%
 }
\caption{$ bits/J $ for $ P_{sleep} = 0 W $ and $ T_{activate} = 5 $}
\label{fig:54}
\end{center}
\end{figure}

\begin{figure}[thb]
\begin{center}
 \centerline{%
 \resizebox{0.7\textwidth}{!}{\includegraphics{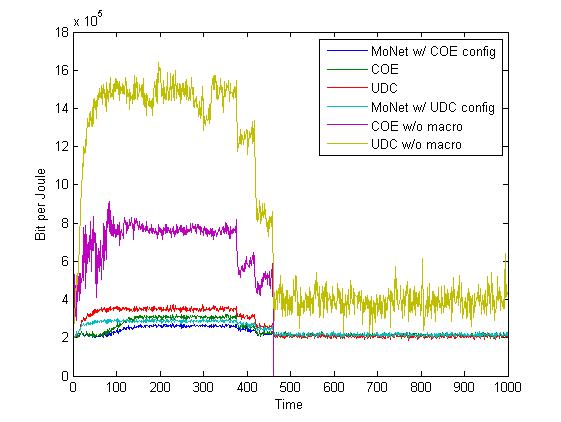}}%
 }
\caption{$ bits/J $ for $ P_{sleep} = 0 W $, $ T_{activate} = 9 $ and $ T_{deactivate} = 4 $}
\label{fig:94}
\end{center}
\end{figure}

\begin{figure}[thb]
\begin{center}
 \centerline{%
 \resizebox{0.7\textwidth}{!}{\includegraphics{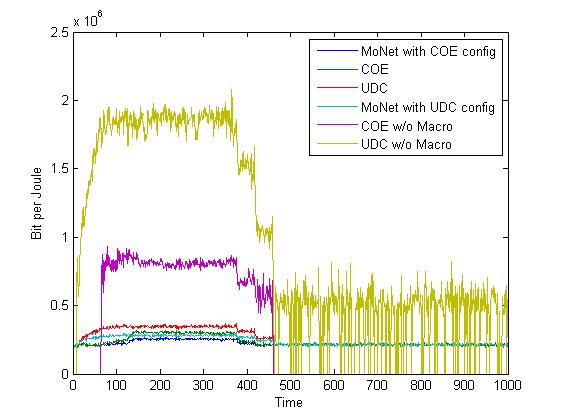}}%
 }
\caption{$ bits/J $ for $ P_{sleep} = 0 W $ and $ T_{activate} = 9 $}
\label{fig:98}
\end{center}
\end{figure}

\begin{figure}[H]
\begin{center}
 \centerline{%
 \resizebox{0.7\textwidth}{!}{\includegraphics{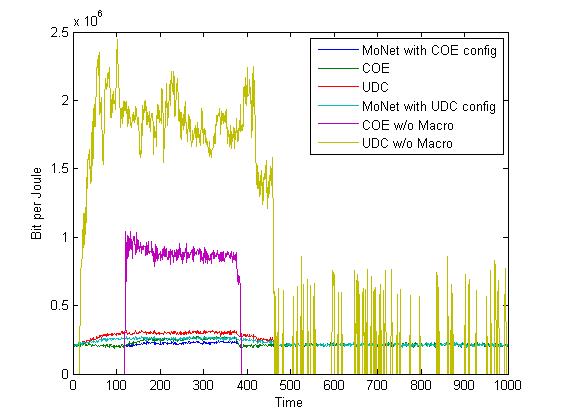}}%
 }
\caption{$ bits/J $ for $ P_{sleep} = 0 W $ and $ T_{activate} = 12 $}
\label{fig:1211}
\end{center}
\end{figure}

In Figure \ref{fig:54}, the average improvement in the $ bits/J $ is around \%20 and even after 470\textsuperscript{th} slot, some pico eNBs in UDC remain active. However, in COE, pico eNBs seldom turn on after 470\textsuperscript{th} slot. At the same time, in Figure \ref{fig:94}, again the average improvement in the $ bits/J $ is around \%20. However, this time, the pico eNBs in COE never remain active after 470\textsuperscript{th} slot. In Figure \ref{fig:98}, once again the average improvement is around \%20. In this case, after 470\textsuperscript{th} slot, some pico eNBs in UDC oscillate between active and sleep modes, as there is only one threshold. But since $ P_{sleep} = 0 W $, this oscillation does not affect the system performance. 

The improvement is around \%20 for Figures \ref{fig:54}, \ref{fig:94} and \ref{fig:98}. This result is expected; since from Figure \ref{fig:sleep0}, we know that the threshold being 5 or 9 do not change the performance of the system. Another observation from Figures \ref{fig:54}, \ref{fig:94} and \ref{fig:98} is that the location of $ T_{deactivate} $ do not affect the system, since between slots 0 and 470, the number of users in pico cells is quite large and the probability of oscillation is almost 0. After 470\textsuperscript{th} slot, even though an oscillation occurs, as $ P_{sleep} = 0 W $, this oscillation does not affect the system performance. On the other hand, in Figure \ref{fig:1211}, the average improvement in the $ bits/J $ is less than \%20, since not all the pico cells are able to turn on, when the threshold is 12. Thus, even though the location of the second threshold does not affect the system performance much, $ T_{activate} $ should be chosen in a way to maximize active pico cells, i.e., it should not be very large.

\begin{figure}[thb]
\begin{center}
 \centerline{%
 \resizebox{0.7\textwidth}{!}{\includegraphics{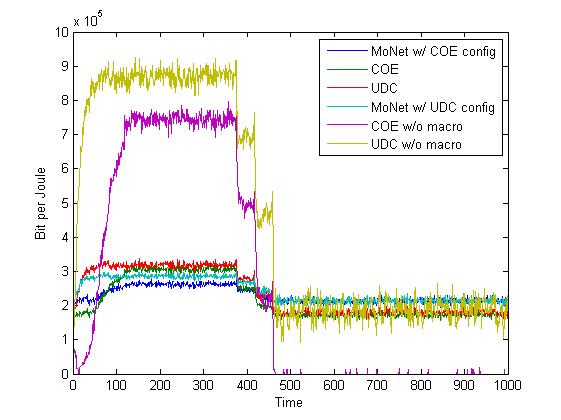}}%
 }
\caption{$ bits/J $ for $ P_{sleep} = 8.6 W $ and $ T_{activate} = 5 $}
\label{fig:548}
\end{center}
\end{figure}

\begin{figure}[thb]
\begin{center}
 \centerline{%
 \resizebox{0.7\textwidth}{!}{\includegraphics{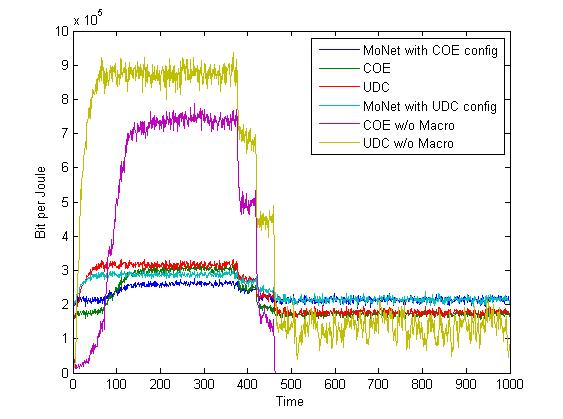}}%
 }
\caption{$ bits/J $ for $ P_{sleep} = 8.6 W $, $ T_{activate} = 9 $ and $ T_{deactivate} = 4 $}
\label{fig:948}
\end{center}
\end{figure}

\begin{figure}[thb]
\begin{center}
 \centerline{%
 \resizebox{0.7\textwidth}{!}{\includegraphics{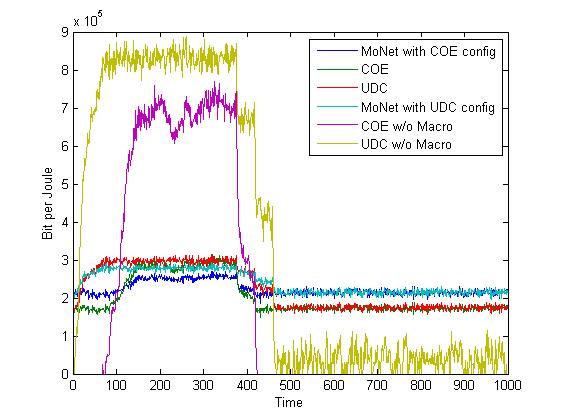}}%
 }
\caption{$ bits/J $ for $ P_{sleep} = 8.6 W $ and $ T_{activate} = 9 $}
\label{fig:988}
\end{center}
\end{figure}

\begin{figure}[H]
\begin{center}
 \centerline{%
 \resizebox{0.7\textwidth}{!}{\includegraphics{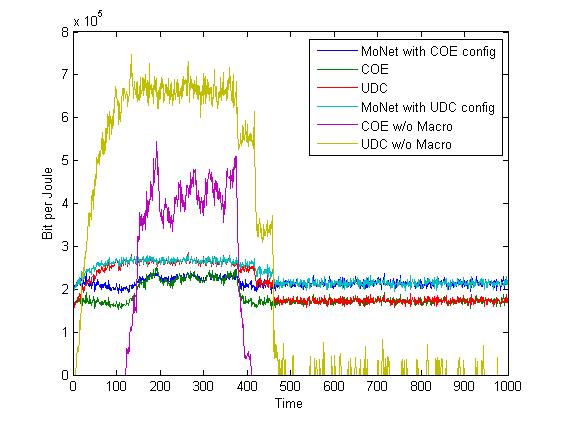}}%
 }
\caption{$ bits/J $ for $ P_{sleep} = 8.6 W $ and $ T_{activate} = 12 $}
\label{fig:12118}
\end{center}
\end{figure}

When $ P_{sleep} = 8.6 W $, the previous graphs transform into Figures \ref{fig:548}, \ref{fig:948}, \ref{fig:988} and \ref{fig:12118}.

In Figures \ref{fig:548}, \ref{fig:948} and \ref{fig:988}, average improvement in the $ bits/J $ between slots 0 and 470 is around \%20, which is the same as before. This is expected since all pico eNBs are active in that interval; thus, the value of $ P_{sleep} $ do not affect the performance of the system between those slots . However, when we compare Figures \ref{fig:948} and \ref{fig:988}, we observe that $ T_{deactivate} $ becomes important in the case where $ P_{sleep} = 8.6 W $. Because, it is more beneficial for the system when more pico cells remain active after 470\textsuperscript{th} slot. Thus, we want  $ T_{deactivate} $ to be as low as possible. We also observe that the performances of COE and UDC are lower than MoNet, after 470\textsuperscript{th} slot. This is because most pico cells are in sleep mode after that slot; thus, in addition to not providing capacity to the system, they consume sleep energy.

Also, when we inspect Figure \ref{fig:12118}, we see that there is no improvement between slots 0 and 470. In addition, after 470\textsuperscript{th} slot, the performance of COE and UDC is worst among all simulations, since almost all pico cells are in sleep mode, both in UDC and COE. In short, in the case of hotspot users, both $ T_{activate} $ and $ T_{deactivate} $ should be chosen as small as possible, since the system benefits most when most pico cells are active.

\section{Conclusion}
\label{chp:conc}
In this paper, a smart sleep strategy for small cell base stations is proposed for
improving the energy efficiency of HetNets. When the number of users within
the coverage area of an inactive small cell base station exceeds the activation
threshold, Ta, the base station is switched to the active mode. On the other
hand, when the number of users within the coverage area of an active small cell
base station drops below the deactivation threshold, $T_d$, the base station is turned
off.
When the users have uniformly distributed initial locations and $P_{sleep = 0}$
for pico eNBs, the choice of $T_a$ is critical. However, when $P_{sleep}>0$, there is a
penalty for turning off an eNB and EE decreases monotonically with $T_a$. Thus,
it is more beneficial for all pico eNBs that have at least one user to be active.
In the case of hotspot users, EE vs. threshold graphs are again monotonically
decreasing, so again, turning on all pico eNBs is more efficient.
On the other hand, when the users' motions are not random, i.e., they move
according to a hotspot model so as to enter and leave pico cells in groups, HetNet
becomes more beneficial. Because, in that model, the number of users per pico
cell increases and in return, the effect of offset power of the base station decreases.
In the case of hotspot users, the bit rate provided by HetNet can get as high as $6.8 10^5$ b/s, which is an increase by $29\%$, compared to MoNet. Meanwhile, the
energy efficiency increases by $20\%$.
In short, using the proposed activity control algorithm, we are able to increase
the average bit rate of the users while consuming the energy more efficiently.
Since, we have simulated various networking scenarios with different topologies,
sleep energies and user distributions, we have been able to analyze the effects of
several parameters on the network performance.
Future research may be directed to consideration of cooperative strategies for
activity control of small cell base stations where joint consideration of number
of users in neighboring small cells is employed. That way, the users within the
coverage of inactive small cell base stations can be served by nearby active small
cell base stations to increase spectral efficiency.

\bibliography{whole}
\bibliographystyle{ieeetr}
\end{document}